\documentclass{aa}
\usepackage{graphics, epsfig, subfigure}
\newcommand{\goodgap}{\hspace{\subfigtopskip} \hspace{\subfigbottomskip}}

\def\lsim{\mathrel{\rlap{\lower3.5pt\hbox{\hskip0.5pt$\sim$}}
    \raise0.5pt\hbox{$<$}}}
\def\gsim{~\rlap{$>$}{\lower 1.0ex\hbox{$\sim$}}}
\def\ML{\mbox{$M/L$}}
\def\Yst{\mbox{$\Upsilon_{*}$}}

\begin{document}

\title{The global mass\,-\,to\,-\,light ratio of SLACS lenses}

\author{V.F. Cardone\inst{1,2,3}, C. Tortora \inst{1,2,4}, R. Molinaro\inst{5},
V. Salzano\inst{1,6}}
\offprints{V.F. Cardone \email{winnyenodrac@gmail.com}}
\institute{Dipartimento di Scienze Fisiche, Universit\`{a} di
Napoli Federico II, Complesso Universitario Monte S. Angelo -
Edificio 6, via Cinthia, 80126 - Napoli, Italy \and Osservatorio
Astrofisico di Catania, via Santa Sofia 78, 95123 - Catania, Italy
\and Dipartimento di Fisica Generale ``A. Avogadro'', Universit\`{a} 
di Torino and Istituto Nazionale di Fisica Nucleare - Sezione di Torino, 
Via Pietro Giuria 1, 10125 - Torino, Italy
\and Osservatorio Astronomico di Capodimonte, Salita Moiariello
16, 80131 - Napoli, Italy \and Dipartimento di Fisica, Politecnico
di Torino, Corso Duca degli Abruzzi 24, 10129 - Torino, Italy \and
Istituto Nazionale di Fisica Nucleare, sezione di Napoli,
Complesso Universitario di Monte S. Angelo - Edificio 6, Via
Cinthia, 80126 - Napoli, Italy}

\date{Received xxx /Accepted yyy}

\abstract{} {The dark matter content of early\,-\,type galaxies
(ETGs) is a hotly debated topic with contrasting results arguing
in favour of or against the presence of significant dark mass within
the effective radius and the change with luminosity and
mass. To address this question, we investigate here the
global mass\,-\,to\,-\,light ratio $\Upsilon(r) = M(r)/L(r)$ of a
sample of 21 lenses observed within the Sloan Lens ACS (SLACS)
survey.} {We follow the usual approach of the galaxy as a
two component systems, but we use a phenomenological ansatz for
$\Upsilon(r)$, proposed by some of us, that is 
able to smoothly interpolate between constant $M/L$ models and a
wide class of dark matter haloes. The resulting galaxy model is
then fitted to the data on the Einstein radius and aperture velocity
dispersion.} {Our phenomenological model turns out to
agree with the data suggesting the presence of massive dark
matter haloes to explain the lensing and dynamics
properties of the SLACS lenses. According to the values of the
dark matter mass fraction, we argue that the halo may play a
significant role in the inner regions probed by the data, but such
a conclusion strongly depends on the adopted initial mass function
of the stellar population. Finally, we find that the dark matter
mass fraction within $R_{eff}$ scales with both the total
luminosity and stellar mass in such a way that more luminous (and
hence more massive) galaxies have a greater dark matter content.}{}


\titlerunning{Global $M/L$ of SLACS lenses}
\authorrunning{V.F. Cardone et al.}

\maketitle

\section{Introduction}

Early\,-\,type galaxies (hereafter ETGs) represent the most
massive and brightest stellar systems in the universe.
Notwithstanding the regularity of their photometric properties and
the existence of remarkable scaling relations, detailed analysis
of their structure has been plagued by both observational and
theoretical shortcomings. Such a frustrating situation mainly stems from
the lack of a reliable mass tracer able to probe the mass
profile outside the effective radius. Although the use
of planetarey nebulae have recently improved the observational
situation allowing the gravitational potential in the
outer regions (\cite{N+01}, 2002, \cite{Romanowsky03}) to be probed, the well
known mass\,-\,anisotropy degeneracy still represent s a
theoretical shortcoming preventing a unique reconstruction of the
mass profile.

Numerical simulations of galaxy formation are usually invoked as a
guide towards understanding the structure of the dark matter (DM)
haloes. Unfortunately, while there is a general consensus on the
mass density profile $\rho_{DM}(r)$ being correctly described by a
double\,-\,power law function with an outer slope
$s = d\ln{\rho}/d\ln{r} = -3$, there is still an open debate concerning
the inner scaling of $\rho_{DM}(r)$. The pioneering result
$\rho_{DM} \sim r^{-1}$ of Navarro, Frenk and White (1996, 1997)
 has been questioned by other simulations providing either a steeper cusp
(e.g., \cite{M98,Ghigna2000,JS00,FM01}) or a shallow profile
(e.g., \cite{P03,N03}). Needless to say, galaxies are two
component systems with the luminous matter playing a not
negligible role in the inner regions which are indeed the most
well probed by the data. While there is general agreement on
the existence and dominance of DM in the outer galaxy regions,
where stars are practically absent (see, e.g., \cite{vdeB03a},
2003b, 2007), paradoxically the DM content in the inner regions is
more difficult to interpret notwithstanding the availability of a
higher number of possible tracers. As pointed out by Mamon \&
Lokas (2005a), the observational data claim for a dominant stellar
component at a radius $ \lsim \, 1 \ R_{eff}$. However, the
uncertainties on which is the stellar initial mass function (IMF),
with the Salpeter (1995) and Chabrier (2001) as leading but not
unique candidates, makes it quite difficult to assess the general
validity of such a result. On large scales, the DM content has
been found to be a strong function of both luminosity and mass
(\cite{Benson2000,MH02,vdeB07}) with a different behaviour between
faint and bright systems. Looking for a similar result for the DM
content within $\sim 1 R_{eff}$ is quite controversial. On one
hand, some authors (\cite{Ger01,BSD2003}) argue for no dependence.
In contrast, other works (e.g.,
\cite{Nap05,Cappellari06,cresc08a}) do find that brighter galaxies
have a larger DM content, while a flattening and a possible
inversion of this trend for lower mass systems, similar to the one
observed for late\,-\,type galaxies (\cite{Persic93}), is still
under analysis (\cite{Nap05,cresc08a}) with no conclusive result
yet obtained.

Given the controversial situation for local galaxies, it is not
surprising that the analysis of high redshift galaxies is still
more difficult considering that only partial information are
usually recovered. For instance, rather than a full velocity
dispersion profile, typically one is able to measure only a single
value describing the global velocity dispersion of the galaxy.
This lack of radial data extent obviously limits the analysis
representing a strong obstacle to any attempts to constrain the
galaxy mass profile. Fortunately, a different mass tracer may be
available for a subset of ETGs at intermediate redshift. Actually,
gravitational lensing gives a strong constraint on the projected
mass within the so called Einstein radius (\cite{SEF,PLW01}). It
is worth stressing that such an observational quantity is fully
model independent although a galaxy model has to be assumed to
extrapolate the results to other radii. Combining strong lensing
with the information from galaxy dynamics makes it possible to not
only increase the number of constraints, but also to partially
break the mass\,-\,anisotropy degeneracy (\cite{TK02,KT03a,B08})
thus representing a powerful tool to probe mass profiles for lens systems.

It is worth noting that most of the analyses in literature heavily
rely on the choice of an analytical expression for the density
profile of the DM halo. Needless to say, this makes the results
model dependent with the risk of biasing them in a difficult to
quantify way. Here, in order to escape this problem, we follow the
approach presented by some of us in Tortora et al. (2007) where a
{\it phenomenological ansatz} for the global \ML\ ratio
$\Upsilon(r) = M(r)/L(r)$ has been proposed. Such a model is able
to smoothly interpolated between constant \ML\ models (where light
traces mass) and the different DM haloes profiles thus providing a
unifying description of a large class of total mass profiles.
Fitting the model to the lensing and dynamics observables makes it
possible to constrain its characterizing parameters thus giving
constraints on the DM mass content and its dependence on the
luminosity or mass of ETGs at intermediate redshift. As a further
step on with respect to our original paper, we here improve the
analysis by first investigating which region of the parameter
space is better suited to mimic the most used dark halo models. We
then fit our model to the observed data on lensing and dynamics in
order to individuate, within the plethora of possible trends
admitted by our parametric choice for \ML\ ratio, the best ones
able to match observations. Such an analysis then makes it also
possible to infer constraints on the DM mass fraction and
investigate whether any dependence on luminosity and stellar mass
can be detected in these intermediate redshift systems.

The paper is organized as follows. In Sect. \ref{sec:model}, we
introduce our parametrization of the global \ML\ ratio and discuss
the main properties of the effective galaxy model coming out from
the combination of our ansatz for $\Upsilon(r)$ and an approximate
analytical deprojection of the Sersic surface brightness profile.
The lensing properties on the model are discussed in Sect.
\ref{sec:GL} where we work out expressions for the deflection and
Einstein angle. The statistical methodology used to contrast the
model against the observations and the dataset are presented in
Sect. \ref{sec:mod_vs_obs}, while a detailed discussion of the
results is the subject of Sect. \ref{sec:results}. We finally
summarize and conclude in Sect. \ref{sec:conclusions}. Along the
paper, where needed, distances are converted from angular to
physical units adopting a flat $\Lambda$CDM cosmological model
with parameters $(\Omega_M, \Omega_{\Lambda}, h) = (0.3, 0.7,
0.7)$ (\cite{WMAP}).

\section{The model}\label{sec:model}

The lack of a reliable mass tracer in the outer regions of ETGs
makes the problem of DM content in these systems still open and
hotly debated. In the usual approach, a dark halo model, described
by a given density profile, is added to the visible component and
then used to fit a given dataset (e.g., the velocity dispersion
profile or the X\,-\,ray emission). Needless to say, such an
approach is strongly model dependent and plagued by parameter
degeneracies so that the constraints on, e.g., the DM mass
fraction within the effective radius $R_{eff}$ are possibly
affected by a theoretical prejudice. In order to overcome such a
problem, some of us have recently advocated the use of a different
strategy. Rather than separately modeling the two galaxy
components, we have proposed to directly constrain its total
(luminous plus dark) cumulative mass profile $M(r)$. Under the
assumption of spherical symmetry, it is possible to write it as\,:

\begin{displaymath}
M(r) = M_{\star}(r) + M_{DM}(r) = \Upsilon(r) L(r)
\end{displaymath}
where $M_{\star}$ ($M_{DM}$) is the mass distribution of the
luminous (dark) component, $L(r)$ the integrated luminosity
profile, and $\Upsilon(r)$ is the {\it global \ML\ ratio}. Since
$L(R)$ may be reconstructed (at least, in principle) by
deprojecting the observed surface brightness, the only unknown
ingredient is just $\Upsilon(r)$. On the one hand, should the DM
content be negligible, which is a borderline case of a
constant \ML\ model, one should expect $\Upsilon(r)$ to remain
approximately constant to the stellar \ML\ value. In contrast,
$\Upsilon(r)$ will be an increasing function of $r$ if the dark
halo plays a significant role since the total mass $M(r)$ should
overcome the stellar one in the outer regions. A unified
description of these two opposite trends is provided by the
following phenomenological ansatz (\cite{p1})\,:

\begin{eqnarray}
\Upsilon(r) & = & \Upsilon_0 \left ( \frac{r}{r_0} \right )^{\alpha}
\left ( 1 + \frac{r}{r_0} \right )^{\beta} \nonumber \\
~ & = & \Upsilon_0 \left ( \frac{\eta}{\eta_0} \right )^{\alpha}
\left ( 1 + \frac{\eta}{\eta_0} \right )^{\beta} \nonumber \\
~ & = & \Upsilon_{eff} \left ( \frac{\eta + \eta_0}{1 + \eta_0} \right
)^{\beta} \eta^{\alpha} \ ,
\label{eq: upsmodel}
\end{eqnarray}
with $\eta = r/R_{eff}$, $\eta_0 = r_0/R_{eff}$ and

\begin{equation}
\Upsilon_{eff} = \frac{\Upsilon_0 (1 + \eta_0)^{\beta}}
{\eta_0^{\alpha + \beta}} \ .
\label{eq: upseff}
\end{equation}
While $(\Upsilon_0, r_0)$ are scaling quantities, the two
parameters $(\alpha, \beta)$ plays a more significant role setting
the asymptotic slopes of the global \ML\ ratio. Indeed, it is\,:

\begin{equation}
\Upsilon(r) \propto \left \{
\begin{array}{ll}
\Upsilon_0 (r/r_0)^{\alpha} & r << r_0 \\
~ & ~ \\
\Upsilon_0 (r/r_0)^{\alpha + \beta} & r >> r_0 \\
\end{array}
\right . \ .
\label{eq: mlslopes}
\end{equation}
Eq.(\ref{eq: mlslopes}) clearly shows that, rather than $\beta$,
it is $\alpha + \beta$ the important quantity so that, in the following,
we will parametrize the model by the parameters $(\alpha, \alpha + \beta,
\eta_0, \Upsilon_{eff})$. It is instructive to look at the following limits\,:

\begin{equation}
\lim_{\eta \rightarrow 0}{\Upsilon(\eta)} =
\left \{
\begin{array}{ll}
0 & \ \ \alpha > 0, \ \alpha + \beta > 0 \\
~ & ~ \\
\infty & \ \ \alpha < 0, \ \alpha + \beta > 0 \\
~ & ~ \\
0 & \ \ \alpha > 0, \ \alpha + \beta < 0 \\
~ & ~ \\
\infty & \ \ \alpha < 0, \ \alpha + \beta < 0 \\
\end{array}
\right . \ ,
\label{eq: limzero}
\end{equation}
while for $\alpha = 0$, we get\,:

\begin{equation}
\lim_{\eta \rightarrow 0}{\Upsilon(\eta)} = \Upsilon_0 \ .
\label{eq: limzeroalphazero}
\end{equation}
According to Eqs.(\ref{eq: limzero}) and (\ref{eq: limzeroalphazero}),
the global \ML\ diverges if $\alpha$ is negative. Considering its
definition and that both $M(r)$ and $L(r)$ identically vanish in the
origin, we shall conclude that, for this class of models, $M(r)$ goes to zero
faster than $L(r)$ which is not expected. Indeed, should visible
mass dominate in the inner regions, we expect $M(r) \simeq
M_{\star}(r) = \Yst\ L(r)$, with \Yst\ the stellar \ML\ ratio,
so that $M(r)$ and $L(r)$ decreases with the same rate.
In the opposite case, i.e. DM is more concentrated than the light,
it is possible to have $M(r)$ decreasing slower than light so that
we cannot discard models with negative $\alpha$. On the other hand,
when $\alpha > 0$, we get a null global \ML\ ratio in the origin which
is an unphysical result since $\Upsilon(r)$ can never become smaller
than the stellar \ML. For this same reason, the case $\alpha = 0$ should
be accepted provided that $\Upsilon_0 \ge \Upsilon_{\star}$ hold.

One could then argue that the case $\alpha > 0$ should be discarded. It is
worth stressing, however, that our ansatz in Eq.(\ref{eq: upsmodel}) is
purely phenomenological. We have looked for a functional shape
that makes it possible to fit as well as possible different realistic
models suitably adjusting the characteristic parameters. As for all
fitting expressions, we do not expect that the fitted approximation
recovers the underlying dark halo model from zero to infinity, but only
over a limited (although large) radial range. As we will see later,
this is what indeed happens with our parametrization of $\Upsilon(r)$
notwithstanding the very inner profile shape. Motivated by these
considerations, we will therefore consider both $\alpha \ge 0$ and $\alpha
< 0$ models.

As an attempt to constrain $(\alpha + \beta)$, we can consider the following limits\,:

\begin{equation}
\lim_{\eta \rightarrow \infty}{\Upsilon(\eta)} =
\left \{
\begin{array}{ll}
\infty & \ \ \alpha \ge 0, \ \alpha + \beta > 0 \\
~ & ~ \\
\infty & \ \ \alpha < 0, \ \alpha + \beta > 0 \\
~ & ~ \\
0 & \ \ \alpha \ge 0, \ \alpha + \beta < 0 \\
~ & ~ \\
0 & \ \ \alpha < 0, \ \alpha + \beta < 0 \\
\end{array}
\right . \ ,
\label{eq: liminfty}
\end{equation}
which show that models with $\alpha + \beta < 0$ are clearly
unphysical. Indeed, in this case, we easily get $M(r \rightarrow
\infty) = 0$, that is to say the total mass of the system is null.
Actually, as we will see later, it is not possible to reproduce
the $\Upsilon(r)$ trend of any DM model with our ansatz (\ref{eq:
upsmodel}) when $\alpha + \beta < 0$ so that we will henceforth
only consider models with nonnegative values of the $\alpha +
\beta$ parameter (\cite{p1}). In this case, the total mass is
divergent, but this is a common feature for many dark halo models
(e.g., the isothermal sphere and the NFW profile). In order to avoid
this problem, one may truncate the halo at the virial radius.

Further hints on the role of the different parameters at work may
be obtained considering the logarithmic slope of the global \ML\
ratio. It is\,:

\begin{equation}
\frac{d\log{\Upsilon(\eta)}}{d\log{\eta}} = \alpha + \beta \eta
(\eta + \eta_0)^{-1}
\label{eq: slopeups}
\end{equation}
so that\,:

\begin{displaymath}
\frac{d\log{\Upsilon(\eta)}}{d\log{\eta}} \ge 0 \iff
\eta \ge \eta_c = - [\alpha/(\alpha + \beta)] \eta_0 \ .
\end{displaymath}
Since we have decided to only consider models with $\alpha + \beta
> 0$, we therefore conclude that the global \ML\ ratio is
everywhere increasing if $\alpha > 0$, while the logarithmic slope
becomes positive only after the threshold radius $\eta_c$ when
$\alpha$ is negative. As we will see, both these trends are
possible for realistic DM models so that we will not set any
constraint on $\eta_c$. It is worth noting that, in a narrow range
around $\eta_c$, the logarithmic slope is approximately null so
that $\Upsilon(\eta)$ is constant as in the
mass\,-\,traces\,-\,light models. Should the data at hand probe
just this radial range, one can reasonably fit them with a
constant \ML\ ratio model.

\subsection{The luminosity density}

The phenomenological ansatz for $\Upsilon(r)$ is just one side of the coin,
the other one being the choice of the luminosity profile. In principle,
this quantity may be directly reconstructed from the data by first
deconvolving the observed surface brightness with the PSF of the images and
then deprojecting it under an assumption for the intrinsic flattening.
Needless to say, such an idealized procedure is strongly affected by noise
and will call for a case\,-\,by\,-\,case study. Fortunately, ETGs are quite
regular in their photometric properties. Indeed, as well known
(\cite{CCD93,GC97,PS97}), their surface brightness is well described by the
Sersic profile (\cite{Sersic})\,:

\begin{equation}
I(R) = I_e \exp{\left \{ - b_n \left [ \left ( \frac{R}{R_{eff}} \right
)^{1/n}
- 1 \right ] \right \}}
\label{eq: ir}
\end{equation}
with $R$ the cylindrical radius\footnote{Note that we have implicitly
assumed that the intensity $I$ does not depend on the angular coordinates.
Actually, the isophotes are not concentric circles, but rather ellipses
with variable ellipticities and position angles so that $I = I(R,
\varphi)$. However, in order to be consistent with our assumption of
spherical symmetry of the three dimensional mass profile, we will neglect
such an effect and, following a common practice, {\it circularize} the
intensity profile considering circular isophothes with radii equal to the
geometric mean of the major and minor axes.} on the plane of the sky and
$I_e$ the luminosity intensity at the effective radius $R_{eff}$. The
constant $b_n$ is determined by the condition that the luminosity within
$R_{eff}$ is half the total luminosity, i.e. $b_n$ is found by solving\,:

\begin{equation}
\Gamma(2n, b_n) = \Gamma(2n)/2
\label{eq: bn}
\end{equation}
where $\Gamma(a, z)$ is the incomplete $\Gamma$ function. Although
Eq.(\ref{eq: bn}) is straightforward to solve numerically, a very good
analytical approximation is given by (\cite{CB99})\,:

\begin{displaymath}
b_n = 2n - \frac{1}{3} - \frac{0.009876}{n} \ .
\end{displaymath}
The deprojection of the intensity profile in Eq.(\ref{eq: ir}) is
straightforward under the hypothesis of spherical symmetry, but,
unfortunately, the result turns out to be a somewhat involved combinations
of the unusual Meijer functions (\cite{MC02}). In order to not deal with
these difficult to handle expression, we prefer to use the model proposed
by Prugniel and Simien (1997, hereafter PS97) whose three dimensional
luminosity density reads\,:

\begin{equation}
j(r) = j_0 \left ( \frac{r}{R_{eff}} \right )^{-p_n} \exp{\left [ -b_n
\left (
\frac{r}{R_{eff}} \right )^{1/n} \right ]}
\label{eq: jr}
\end{equation}
with

\begin{equation}
j_0 = \frac{I_0 b_n^{n (1 - p_n)}}{2 R_{eff}} \frac{\Gamma(2n)}{\Gamma[n (3
- p_n)]} \ .
\label{eq: jz}
\end{equation}
Here, $I_0 = I(R = 0) = I_e {\rm e}^{b_n}$, while the constant $p_n$ is
chosen so that the projection of Eq.(\ref{eq: jr}) matches a Sersic profile
with the same values of $(n, R_{eff}, I_e)$. A useful fitting formula is
given as (\cite{Metal01})\,:

\begin{displaymath}
p_n = 1.0 - \frac{0.6097}{n} + \frac{0.00563}{n^2} \ .
\end{displaymath}
Because of the assumed spherical symmetry, the luminosity profile may
simply be obtained as\,:

\begin{displaymath}
L(r) = 4 \pi \int_{0}^{r}{r'^2 j(r') dr'}
\end{displaymath}
which, for the PS model, becomes\,:

\begin{equation}
L(r) = L_T {\times} \frac{\gamma[n (3 - p_n), b_n \eta^{1/n})}{\Gamma[n (3
- p_n)]}
\label{eq: lr}
\end{equation}
where the total luminosity $L_T$ reads\,:

\begin{equation}
L_T = 2 \pi n b_n^{-2n} {\rm e}^{b_n} \Gamma(2n) I_e R_{eff}^2 \ .
\label{eq: lt}
\end{equation}
Note that the total luminosity is the same as the projected one for the
corresponding Sersic profile as can be immediately check computing\,:

\begin{displaymath}
L_T = 2 \pi \int_{0}^{\infty}{I(R) R dR} \ .
\end{displaymath}
It is worth remembering the definitions of the different special functions
entering Eqs.(\ref{eq: bn})\,-\,(\ref{eq: lt})\,:

\begin{displaymath}
\Gamma(a) = \int_{0}^{\infty}{t^{a - 1} {\rm e}^{-t} dt} \ ,
\end{displaymath}

\begin{displaymath}
\Gamma(a, y) = \int_{y}^{\infty}{t^{a - 1} {\rm e}^{-t} dt} \ ,
\end{displaymath}

\begin{displaymath}
\gamma(a, y) = \int_{0}^{y}{t^{a - 1} {\rm e}^{-t} dt} \ ,
\end{displaymath}
so that the following useful relations hold\,:

\begin{displaymath}
\Gamma(a) = \lim_{y \rightarrow 0}{\Gamma(a, y)} = \lim_{y \rightarrow
\infty}{\gamma(a, y)} \ ,
\end{displaymath}

\begin{displaymath}
\gamma(a, y) = \Gamma(a) - \Gamma(a, y) \ .
\end{displaymath}
As a final remark, let us stress that the total mass of the
stellar component $M_{\star}^T = \Yst\ L_T$ may be determined from
the measurement of the photometric parameters $(n, R_{eff}, I_e)$
provided that an estimate of the stellar \ML\ ratio \Yst\ is
available (for instance, from the relation between \Yst\ and the
colours or from fitting the galaxy spectrum to stellar population
synthesis models).

\subsection{The total mass, density profile and DM fraction}

Combining Eq.(\ref{eq: upsmodel}) and (\ref{eq: lr}), we trivially get for
the total mass profile\,:

\begin{eqnarray}
M(r) & = & \Upsilon_0 L_T \left ( \frac{\eta}{\eta_0} \right )^{\alpha}
\left ( 1 + \frac{\eta}{\eta_0} \right )^{\beta} \frac{\gamma[n(3 - p_n), b_n
\eta^{1/n}]}{\Gamma[n(3 - p_n)]} \nonumber \\ ~ & = & \Upsilon_{eff}
L_T \ \frac{\eta^{\alpha} (\eta + \eta_0)^{\beta} \gamma[n(3 - p_n), b_n \eta^{1/n}}
{(1 + \eta_0)^{\beta} \ \Gamma[n(3 - p_n)]}.
\label{eq: mr}
\end{eqnarray}
Formally, $M(r)$ diverges if $\alpha + \beta > 0$. However, this is not a
serious problem since, following a common practice, we will truncate the
model at the virial radius $R_{vir}$ defined as the radius of the sphere
within which the mean density is $\Delta_c(z) \bar{\rho}_M(z)$ with\,:

\begin{displaymath}
\bar{\rho}_M(z) = \Omega_{0M} \rho_{crit} (1 + z)^3 = (3H_0^2/8 \pi G) \Omega_{0M} (1 + z)^3 \ ,
\end{displaymath}
and, following Bryan \& Norman (1998), we approximate the virial overdensity as\,:

\begin{displaymath}
\Delta_c(z) = \frac{18 \pi + 82 x(z) - 39 x^2(z)}{\Omega_{0M}}
\end{displaymath}
where $\Omega_{0M}$ is the present day matter density parameter and, for the flat $\Lambda$CDM
model we are considering, it is\,:

\begin{displaymath}
x(z) = \Omega_M(z) - 1 = \frac{\Omega_{0M} (1 + z)^3}
{\sqrt{\Omega_{0M} (1 + z)^3 - (1 - \Omega_{0M})}} - 1 \ .
\end{displaymath}
The total mass is then the virial one obtained
by setting $\eta = \eta_{vir} = R_{vir}/R_{eff}$ in Eq.(\ref{eq: mr}) with\,:

\begin{equation}
R_{vir} = \left [ \frac{3 M_{vir}}{4 \pi \Delta_c
\bar{\rho}_M(z)} \right ]^{1/3} \ ,
\end{equation}
and solving with respect to $M_{vir}$. This gives\,:

\begin{displaymath}
M_{vir} = \Upsilon_{eff} L_T \ \times \
\frac{\eta_{vir}^{\alpha} (\eta_{vir} + \eta_0)^{\beta}
\gamma[n(3 - p_n), b_n \eta^{1/n}]}{(1 + \eta_0)^{\beta} \ \Gamma[n(3 - p_n)]} \ ,
\end{displaymath}
so that one can conveniently rewrite the mass profile as\,:

\begin{eqnarray}
M(r) & = & M_{vir} \nonumber \\
~ & \times & \frac{(\eta/\eta_0)^{\alpha} (1 + \eta/\eta_0)^{\beta}
\gamma[n(3 - p_n), b_n \eta^{1/n}]}{(\eta_{vir}/\eta_0)^{\alpha} (1 +
\eta_{vir}/\eta_0)^{\beta} \gamma[n(3 - p_n), b_n \eta_{vir}^{1/n}]}
\label{eq: mrbis}
\end{eqnarray}
which clearly shows that $(R_{eff}, I_e, \Upsilon_{eff})$ only
works as scaling parameters through $M_{vir}$, while the shape of
$M(r)$ is controlled by the Sersic index $n$ and the \ML\ slopes
$(\alpha, \beta)$. Note that the {\it light traces mass} case is
only achieved by setting both $\alpha = \beta = 0$ and
$\Upsilon_{eff} = \Yst$, with \Yst\ the stellar \ML\ ratio.
Varying these parameters, Eq.(\ref{eq: mrbis}) provides a wide
range of possible mass profiles thus allowing a rather
large class of dark halo models to be mimicked. Note also that, in principle, one
should also check that $M(r)$ is a monotonically increasing
function of $r$. To this end, it is convenient to consider the
logarithmic slope of the mass profile reading\,:

\begin{eqnarray}
\frac{d\ln{M}}{d\ln{\eta}} & = & \frac{\left ( b_n \eta^{1/n}
\right )^{n(3 - p_n)} {\rm e}^{-b_n \eta^{1/n}}}{n \gamma[n(3 - p_n),
b_n \eta^{1/n}]} \nonumber \\
~ & + & \frac{\left [ \alpha (\eta + \eta_0) + \beta \eta \right ]
\gamma[n(3 - p_n), b_n \eta^{1/n}]}{\eta + \eta_0} \ .
\end{eqnarray}
It is instructive to first consider the case $n = 4$. We get\,:

\begin{displaymath}
\lim_{\eta \rightarrow 0}{\frac{d\ln{M}}{d\ln{\eta}}} = \alpha \ \
, \ \ \lim_{\eta \rightarrow \infty}{\frac{d\ln{M}}{d\ln{\eta}}} =
\alpha + \beta \ \ ,
\end{displaymath}
so that only models with $\alpha \ge 0$ and $\alpha + \beta \ge 0$ are viable.
Leaving $n$ as a free parameter, we find\,:

\begin{displaymath}
\frac{d\ln{M}}{d\ln{\eta}} \ge 0 \iff \eta \ge \eta_c =
- \frac{\alpha \eta_0}{\alpha + \beta} \ .
\end{displaymath}
Not surprisingly, this is the same constraint we have obtained
when considering the logarithmic slope of the global \ML\ ratio
$\Upsilon(r)$ so that we refer the reader to the above discussion.
We only stress here that models with $\alpha < 0$ have an
unphysical mass profile in the inner regions. As such, the behaviour
of these models for such low values of $r$ should not be considered
as a good approximation of the reality. Indeed, we get\,:

\begin{displaymath}
\lim_{\eta \rightarrow 0}{M(\eta, \alpha < 0)} = \pm \infty \ .
\end{displaymath}
which is clearly an unphysical result. However, one can consider
our ans\"atz for $\Upsilon(r)$ only as a phenomenological approximation
working well everywhere but for $\eta \le \eta_c$. Since this quantity
is typically quite small, hereafter, we will consider both negative and
positive values of $\alpha$, while only models with $\alpha + \beta > 0$
will be explored to avoid dealing with an unphysical vanishing total mass.

Because of the assumed spherical symmetry, the mass density may be
straightforwardly derived as\,:

\begin{displaymath}
\rho(r) = \frac{1}{4 \pi r^2} \frac{dM}{dr} = \frac{1}{4 \pi R_{eff}^3} \frac{1}{\eta^2}
\frac{dM}{d\eta} \ .
\end{displaymath}
Using Eq.(\ref{eq: mrbis}) and some lengthy algebra, we finally get\,:

\begin{equation}
\rho(\eta) = \frac{M_{vir}}{4 \pi R_{eff}^3 {\cal{N}}_{vir}}
\tilde{\rho}_ 1(\eta) \tilde{\rho}_2(\eta)
\label{eq: rhovseta}
\end{equation}
with\,:

\begin{equation}
{\cal{N}}_{vir} = \left ( \frac{\eta_{vir}}{\eta_0} \right )^{\alpha}
\left ( 1 + \frac{\eta_{vir}}{\eta_0} \right )^{\beta}
\gamma[n(3 - p_n), b_n
\eta_{vir}^{1/n}] \ ,
\label{eq: defnvir}
\end{equation}

\begin{equation}
\tilde{\rho}_1 = \left ( \frac{\eta}{\eta_0} \right )^{\alpha - 3}
\left ( 1 + \frac{\eta}{\eta_0} \right )^{\beta - 1} \exp{(-b_n \eta^{1/n})} \ ,
\label{eq: defrho1}
\end{equation}

\begin{eqnarray}
\tilde{\rho}_2 & = & \left ( 1 + \frac{\eta}{\eta_0} \right )
(b_n \eta^{1/n})^{n(3 - p_n)}
\nonumber \\ ~ & + & n \alpha \left [ 1 +
\left ( \frac{\alpha + \beta}{\alpha} \right ) \eta \right ]
\gamma[n(3 - p_n), b_n \eta^{1/n}] \ {\rm e}^{b_n \eta^{1/n}} \ .
\label{eq: defrho2}
\end{eqnarray}
It is worth stressing that Eq.(\ref{eq: rhovseta}) reduces to the
stellar mass profile (\ref{eq: jr}) setting $\alpha = \beta = 0$
and $\Upsilon_{eff} = \Yst$ as expected, while it is positive everywhere
for all other values of the parameters provided the above
constraints on $\alpha$ and $\alpha + \beta$ are set. In order to
get further insights on the shape of the density profile, one may
consider its logarithmic slope\,:

\begin{displaymath}
s(\eta) = \frac{d\ln{\rho}}{d\ln{r}} =
\frac{\eta}{\rho} \frac{d\rho}{d\eta} \ .
\end{displaymath}
Some lengthy algebra finally gives\,:

\begin{eqnarray}
s(\eta) = s_1(\eta) + s_2(\eta) \ ,
\label{eq: seta}
\end{eqnarray}
having defined\,:

\begin{equation}
s_1(\eta) = \frac{(\alpha - 3) + (\alpha + \beta - 4)(\eta/\eta_0)}
{1 + \eta/\eta_0} - (b_n/n) \eta^{1/n} \ ,
\label{eq: s1}
\end{equation}

\begin{equation}
s_2(\eta) = \frac{\eta {\cal{N}}_2(\eta)}{\eta_0 {\cal{D}}_2(\eta)} \ ,
\label{eq: s2}
\end{equation}

\begin{eqnarray}
{\cal{N}}_2 & = & (b_n \eta^{1/n})^{n(3 - p_n)} +
b_n^{n(3 - p_n)} \eta_0 \eta^{2 - p_n} \nonumber \\
~ & \times & \left \{ (3 - p_n)(1 + \eta/\eta_0)
+ \alpha \left [ 1 + (\alpha + \beta)(\eta/\eta_0)/\alpha \right ] \right \}
\nonumber \\ ~ & + & \left \{ n(\alpha + \beta) + \alpha \eta_0
\left [ 1 + (\alpha + \beta)(\eta/\eta_0)/\alpha \right ]
\eta^{-1 + 1/n} \right \} \nonumber \\ ~ & \times &
\gamma[n(3 - p_n), b_n \eta^{1/n}] \ ,
\label{eq: n2}
\end{eqnarray}

\begin{eqnarray}
{\cal{D}}_2 & = & (b_n \eta^{1/n})^{n(3 - p_n)} (1 + \eta/\eta_0)
+ n \alpha \gamma[n(3 - p_n), b_n \eta^{1/n}] \nonumber \\
~ & \times & \left [ 1 + (\alpha + \beta)(\eta + \eta_0)/\alpha \right ]
\exp{(b_n \eta^{1/n})}  \ .
\label{eq: d2}
\end{eqnarray}
The density profile may be locally approximated as a power law
with a running slope, i.e. we may write $\rho(\eta) \sim
\eta^{s(\eta)}$. It is therefore particularly interesting to
consider the asymptotic limits of the logarithmic slope. To this aim,
we first set $n = 4$ to get\,:

\begin{equation}
\lim_{\eta \rightarrow 0}{s(\eta, n = 4)} = \alpha - 3 \ ,
\label{eq: sin}
\end{equation}

\begin{equation}
\lim_{\eta \rightarrow \infty}{s(\eta, n = 4)} = -\infty
\label{eq: sout}
\end{equation}
so that we get the constraint $\alpha \le 3$ in order that
$\rho(\eta)$ is not an increasing function of the radius in the
inner regions. Note also that models with $\alpha = 3$ provide a
density profile with an inner core thus resembling popular cases
such as the non singular isothermal sphere. On the other hand, in
the asymptotic limit $\eta \rightarrow \infty$, the density
profile exponentially fades away. Such a behaviour should not be
surprising being trivially a consequence of the exponential decay
of the luminosity density profile $j(r)$. Although somewhat
extreme, such a shape in the outer regions is not unrealistic.
Indeed, recent high resolution N\,-\,body simulations
(\cite{N03,Mer06}) show that the global mass profile is well
approximated by an Einasto model (\cite{E65,EH89,PoLLS}) which
indeed predicts such an exponential decay\footnote{Actually, these
are DM only simulations so that the addition of a baryonic
component changes the global mass profile. Nevertheless, in the
very outer regions, the stellar component may be neglected so that
it is likely that the global mass profile behaves as the DM one
thus recovering the exponential decay predicted by these
simulations.}. It is worth stressing that the same results are
obtained also for other values of $n$ as we have checked
numerically (for $n \ge 2$) so that the constraint on $\alpha$ can
be used in the following analysis.

As a final cautionary remark, we stress that the logarithmic slope
$s(\eta)$ in Eq.(\ref{eq: seta}) refers to the density law of the
effective model defined by our ansatz for the global \ML\ ratio.
As such, $s(\eta)$ is not the logarithmic slope of the (eventually
present) dark halo model. Actually, while in the outer regions the
stellar density may be neglected so that $s(\eta)$ may be
identified with the $s_{DM}(\eta)$, in the inner regions things
are more involved. In particular, an inner core of the effective
density profile, i.e. $s(\eta \rightarrow 0) = 0$, does not imply
that the corresponding dark halo model is a cored one.

Finally, one of the main outcomes of our analysis is the spherical
DM content in the core of galaxies, typically within $1 \ R_{eff}$.
We define the spherical DM fraction within the radius $r$ as

\begin{equation}
f_{DM}(r) = \frac{M_{DM}(r)}{M_{DM}(r)+M_{\star}(r)} = 1 -
\frac{M_{\star}(r)}{M(r)}
\end{equation}
where $M_{\star}(r)=\Yst L(r)$ and $M_{DM}(r)$ is DM mass.

\subsection{The surface mass density}

A further quantity to be discussed is the surface mass density,
which we compute as\,:

\begin{displaymath}
\Sigma(R) = 2 \int_{0}^{\infty}{\rho(r) dz}
\end{displaymath}
having defined $R = (x^2 + y^2)^{1/2}$ as usual. It is
convenient to rewrite the above integral as\,:

\begin{displaymath}
\Sigma(R) = 2 \int_{0}^{\pi/2}{\frac{\rho(R/\sin{\theta})
R}{\sin^2{\theta}} d\theta}
\end{displaymath}
with $r = R/\sin{\theta}$, so that, inserting Eq.(\ref{eq: rhovseta})
for $\rho(r)$ and defining $\xi = R/R_{eff}$, after some algebra, we
finally get\,:

\begin{equation}
\Sigma(\xi) =  \frac{M_{vir} \ \xi^{\alpha - 2}}{2 \pi n
\eta_0^{\alpha} R_{eff}^2 {\cal{N}}_{vir}} \ \tilde{\Sigma}(\xi) =
\frac{\Sigma_{eff}}{\xi^{2 - \alpha}}\
\frac{\tilde{\Sigma}(\xi)}{\tilde{\Sigma}(\xi = 1)} \label{eq:
sigmamass}
\end{equation}
with\,:

\begin{eqnarray}
\tilde{\Sigma}(\xi) & = & \int_{0}^{\pi/2}{\left (1 +
\frac{\xi/\eta_0}{\sin{\theta}} \right )^{\beta - 1}
\frac{\tilde{\rho}_2(\xi/\sin{\theta})}{\sin^{\alpha - 1}{\theta}}
} \nonumber \\ ~ & {\times} & \exp{\left [ -b_n \left (
\frac{\xi}{\sin{\theta}} \right )^{1/n} \right ]} d\theta \ ,
\label{eq: defsigmatilde}
\end{eqnarray}

\begin{eqnarray}
\Sigma_{eff} & = & \frac{M_{vir}}{2 \pi R_{eff}^2} \
\frac{\tilde{\Sigma}(\xi = 1)}{n \eta_0^{\alpha} {\cal{N}}_{vir}}
\nonumber \\ ~ & = & \frac{\Upsilon_0 L_{T}}{2 \pi R_{eff}^2} \
\frac{\tilde{\Sigma}(\xi = 1)} {n \eta_0^{\alpha} \Gamma[n(3 -
p_n)]} \nonumber \\ ~ & = & \frac{\Upsilon_{eff} L_T}{2 \pi
R_{eff}^2} \ \frac{\eta_0^{\beta} \tilde{\Sigma}(\xi = 1)} {n (1 +
\eta_0)^{\beta} \Gamma[n(3 - p_n)]} \ . \label{eq: defsigmae}
\end{eqnarray}
Eq.(\ref{eq: sigmamass}) clearly shows the role of the different
model parameters. Indeed, while $(R_{eff}, I_e, \Upsilon_{eff})$
only work as scaling quantities through $\Sigma_{eff}$, the shape
of the surface mass density profile is governed by the Sersic
index $n$, the slope parameters $(\alpha, \beta)$ and the
transition radius $\eta_0$. Ideally, one can therefore put further
constraints on $(\alpha, \beta)$ demanding that $\Sigma(\xi)$ is a
decreasing function in order to be physically viable.
Unfortunately, such a program is not feasible. As Eq.(\ref{eq:
sigmamass}) shows, there are two terms determining the shape of
$\Sigma(\xi)$. While the first one scales as $\xi^{\alpha - 2}$,
it is not possible to infer a general scaling rule for
$\tilde{\Sigma}(\xi)$ since it has to be evaluated numerically. We
have therefore investigated the shape of $\Sigma(\xi)$ by
evaluating it numerically for different values of $(n, \alpha,
\alpha + \beta, \eta_0)$. It turns out that the surface density is
a monotonically decreasing function for $\xi > 0.01$ whatever are
the values of the model parameters. In the very inner region,
however, it is sometimes possible to get a cored profile or an
unphysical increasing $\Sigma(\xi)$. Since we are unable to work
out any analytical rule to identify these models, we prefer to not
put any further constraints on the model parameters. Note,
however, that, also in this disturbing circumstances,
$\Sigma(\xi)$ is bad behaved only in a very inner region where our
phenomenological ans\"atz may break down. Nevertheless, as we will
see later, this has no effect on the estimate of the quantities of
interest.

\subsection{Projected mass and DM fraction}

A quite interesting derived quantity, which will be useful when
contrasting models against observations, is the projected mass
within a circular aperture of radius $R$ given by\,:

\begin{displaymath}
M_{proj}(R) = 2 \pi \int_{0}^{R}{\Sigma(R') R' dR'} = 2 \pi
R_{eff}^2 \int_{0}^{\xi}{\Sigma(\xi') \xi' d\xi'}
\end{displaymath}
which, for our model, reduces to\,:

\begin{equation}
M_{proj}^{th}(\xi) = \frac{\Upsilon_{eff} L_T}{n \Gamma[n(3 -
p_n)]} \left ( \frac{\eta_0}{1 + \eta_0} \right )^{\beta}
\int_{0}^{\xi}{\tilde{\Sigma}(\xi') \xi'^{\alpha - 1} d\xi'} \ .
\label{eq: massprojth}
\end{equation}
For the Prugniel\,-\,Simien model we are adopting for the
luminosity component, the projected mass is given by\,:

\begin{equation}
M_{proj}^{PS}(\xi) = \Yst\ L_T \left [ 1 - \frac{\Gamma(2n, b_n
\xi^{1/n})}{\Gamma(2n)} \right ] \label{eq: massprojps}
\end{equation}
so that the projected DM mass fraction reads\,:

\begin{eqnarray}
f_{proj}(\xi) & = & 1 -
\frac{M_{proj}^{PS}(\xi)}{M_{proj}^{th}(\xi)} \nonumber \\ ~ & = &
1 - \frac{n \Gamma[n(3 - p_n)]}{\Upsilon_{eff}/\Yst}
\left ( \frac{1 + \eta_0}{\eta_0} \right )^{\beta} \nonumber \\
~ & ~ & \ \ \ \times \ \frac{1 - \Gamma(2n, b_n
\xi^{1/n})/\Gamma(2n)} {\int_{0}^{\xi}{\tilde{\Sigma}(\xi')
\xi'^{\alpha - 1} d\xi'}} \ . \label{eq: fproj}
\end{eqnarray}
It is worth noting that the projected mass fraction is typically
higher than the spherical mass fraction $f_{DM}(r)$. This is a
consequence of the integration along the line of sight which is
performed in order to evaluate $M_{proj}(R)$. Indeed, the stellar
component contributes mainly to the inner region, while the DM
halo extends over a larger part of the integration domain. As a
result, $f_{proj}(r)$ is higher than $f_{DM}(r)$, with the ratio
$f_{DM}(r)/f_{proj}(r)$ depending both on $r$ and the
concentration of both the stars and dark halo density profiles.

\section{Dynamical and gravitational lensing properties}\label{sec:GL}

While photometric observations probe the light distribution (hence
giving constraints on the luminosity density), dynamical observables
(e.g., velocity dispersion) and gravitational lensing are excellent probes
of the mass profile. In the framework of our phenomenological scenario,
we can tell that dynamical and lensing observables make it possible to
constrain the global \ML\ ratio $\Upsilon(r)$. Therefore, it is
interesting to derive these main properties for our model.

\subsection{Velocity dispersion}

A widely used probe to constrain the model parameters is
represented by the line of sight velocity dispersion luminosity
weighted within a circular aperture of radius $R_{ap}$. This can
be easily evaluated as\,:

\begin{displaymath}
\sigma_{ap}^2 = \frac{\int_{0}^{R_{ap}}{I(R) \sigma_{los}^2(R) R
dR}} {\int_{0}^{R_{ap}}{I(R) R dR}}
\end{displaymath}
with $\sigma_{los}(R)$ the velocity dispersion projected along the
line of sight. This latter may be found solving the Jeans equation
provided an expression for the anisotropy parameter is chosen. In
order to not further increase the number of unknown quantities in
the problem, we assume isotropy in the velocity space. In this
case, it is possible to arrive at the following expression for
$\sigma_{ap}$ (\cite{MamonLokas05a})\,:

\begin{eqnarray}
\sigma_{ap}^2 & = & \frac{4 \pi G}{3 L_2(R_{ap})} \left [
\int_{0}^{\infty}{r j(r) M(r) dr} \right . \nonumber \\ ~ & - &
\left . \int_{R_{ap}}^{\infty}{(r^2 - R_{ap}^2)^{3/2} j(r) M(r)
r^{-2} dr} \right ] \ , \label{eq: sigmazero}
\end{eqnarray}
with
\begin{equation}
L_2(R) = \int_{0}^{R}{I(R') R' dR'} \label{eq: defl2}
\end{equation}
the projected luminosity profile. For the model we are
considering, after some algebra, we finally get\,:

\begin{eqnarray}
\sigma_{ap,th}^2 & = & \frac{G \Upsilon_0 L_T}{R_{eff}} \
\frac{b_n^{n(3 - p_n)} \Gamma(2n)}{3 n \Gamma[n(3 - p_n)]} \
\frac{{\cal{I}}_{ap}(\xi_{ap}, {\bf p})}{\gamma(2n, b_n
\xi_{ap}^{1/n})} \nonumber \\ ~ & = & \frac{G \Upsilon_{eff}
L_T}{R_{eff}} \ \frac{b_n^{n(3 - p_n)} \Gamma(2n)}{3 n \Gamma[n(3
- p_n)]} \
\left ( \frac{\eta_0^\alpha}{1 + \eta_0} \right )^{\beta} \nonumber \\
~ & \times & \frac{{\cal{I}}_{ap}(\xi_{ap}, {\bf p})} {\gamma(2n,
b_n \xi_{ap}^{1/n})} \ , \label{eq: sigmazerous}
\end{eqnarray}
where we set $\xi_{ap} = R_{ap}/R_{eff}$, being $R_{ap} = 1.5 \
\rm arcsec$ the aperture radius (equal for all lenses because
$\sigma_{ap}$ is measured with the same instrumental setup) and
have defined\,:

\begin{equation}
{\cal{I}}_{ap} = {\cal{I}}_{ap}^{(1)} - {\cal{I}}_{ap}^{(2)} \ ,
\label{eq: defiap}
\end{equation}

\begin{eqnarray}
{\cal{I}}_{ap}^{(1)} & = & \int_{0}^{\infty}{\left (
\frac{\eta}{\eta_0} \right )^{\alpha} \left (1 +
\frac{\eta}{\eta_0} \right )^{\beta} \left \{ 1 - \frac{\Gamma[n(3
- p_n), b_n \eta^{1/n}]}{\Gamma[n(3 - p_n)]} \right \}} \nonumber
\\ ~ & {\times} & \exp{(-b_n \eta^{1/n})} \eta^{1 - p_n} d\eta \ ,
\label{eq: defiap1}
\end{eqnarray}

\begin{eqnarray}
{\cal{I}}_{ap}^{(2)} & = & \int_{\xi_{ap}}^{\infty}{\left (
\frac{\eta}{\eta_0} \right )^{\alpha} \left (1 +
\frac{\eta}{\eta_0} \right )^{\beta} \left \{ 1 - \frac{\Gamma[n(3
- p_n), b_n \eta^{1/n}]}{\Gamma[n(3 - p_n)]} \right \}} \nonumber
\\ ~ & {\times} & \exp{(-b_n \eta^{1/n})} (\eta^2 -
\xi_{ap}^2)^{3/2} \eta^{-(2 + p_n)} d\eta \ . \label{eq: defiap2}
\end{eqnarray}
For given values of the model parameters, Eq.(\ref{eq: sigmazerous})
allows us to predict the value of $\sigma_{ap}$. However, because of
the uncertaintits on both the photometric parameters and the
stellar $M/L$ ratio, the predicted $\sigma_{ap}$ will be affected by
an uncertainty which we compute by the simple propagation of errors.
We do not report here the final expression for sake of shortness,
but caution  the reader that the uncertainty on $R_{eff}$ must
be propagated also in the denominator term $\gamma(2n, b_n
\xi_{ap}^{1/n})$, while we neglect the contribution to the
integral for simplicity.

\subsection{Lensing observables}

Gravitational lensing is a powerful tool to further constrain
the space of parameters. After having determined the surface mass
density in the previous section, here, we define the deflection
angle, that allows a model independent estimate of
projected mass at Einstein radius to be determined.

\subsubsection{The deflection angle}

A key role in the determination of the lensing properties of a given model
is played by the deflection angle. For a spherically symmetric lens, this
reads (\cite{SEF})\,:

\begin{displaymath}
\hat{\alpha} = \frac{2}{R} \int_{0}^{R}{\frac{\Sigma(R')}{\Sigma_{crit}} R' dR'}
\end{displaymath}
where the critical surface density $\Sigma_{crit}$ is\,:

\begin{displaymath}
\Sigma_{crit} = \frac{c^2 D_s}{4 \pi G D_L D_{LS}}
\end{displaymath}
with $D_S$, $D_L$, $D_{LS}$ the observer\,-\,source,
observer\,-\,lens and lens\,-\,source comoving angular diameter distances, respectively.
Inserting Eq.(\ref{eq: sigmamass}), the deflection angle for our model
is easily computed as\,:

\begin{eqnarray}
\hat{\alpha}(\xi) & = & \frac{2 R_{eff}}{\tilde{\Sigma}(\xi = 1)} \ \frac{\Sigma_{eff}}{\Sigma_{crit}}
\ \frac{1}{\xi} \ \int_{0}^{\xi}{\tilde{\Sigma}(\xi') \xi'^{\alpha - 1} d\xi'} \nonumber \\
~ & = & \frac{\hat{\alpha}_{eff}}{\xi} \ \frac{\tilde{\alpha}(\xi)}{\tilde{\alpha}(\xi = 1)}
\label{eq: defangle}
\end{eqnarray}
where we have defined\footnote{A dimensional remark is in order
here. The deflection angle is expressed in $arcsec$ so that,
although the effective radius $R_{eff}$ is a length, in the right
hand side of Eq.(\ref{eq: defangle}), it must be expressed in
angular units.}\,:

\begin{equation}
\tilde{\alpha}(\xi) = \int_{0}^{\xi}{\xi'^{\alpha - 1} \tilde{\Sigma}(\xi') d\xi'} \ ,
\label{eq: defalphatilde}
\end{equation}

\begin{eqnarray}
\hat{\alpha}_{eff} & = & 2 R_{eff} \ \frac{\Sigma_{eff}}{\Sigma_{crit}} \
\frac{\tilde{\alpha}(\xi = 1)}{\tilde{\Sigma}(\xi = 1)} \nonumber \\
~ & = & 2 R_{eff} \frac{\Upsilon_0 L_T}{2 \pi \Sigma_{crit} R_{eff}^2} \
\frac{\tilde{\alpha}(\xi = 1)}{n \eta_0^{\alpha} \Gamma[n(3 - p_n)]} \nonumber \\
~ & = & 2 R_{eff} \frac{\Upsilon_{eff} L_T}{2 \pi \Sigma_{crit} R_{eff}^2} \
\frac{\eta_0^{\beta} \tilde{\alpha}(\xi = 1)}{n (1 + \eta_0)^{\beta} \Gamma[n(3 - p_n)]} \ .
\label{eq: defalphae}
\end{eqnarray}
Looking at Eqs.(\ref{eq: defangle}) and (\ref{eq: defalphae}), it
is immediately clear that the Sersic index $n$ and the model
parameters $(\alpha, \beta, \eta_0)$ determine the shape of the
deflection angle, while the effective radius $R_{eff}$, the total
luminosity $L_T$ and the mass parameter $\Upsilon_{eff}$ only work
as scaling quantities.

\subsubsection{The Einstein angle}

Should the source be a distant pointlike object (such as a
quasar), one may observe multiple images and then use their
positions to constrain the model parameters. On the other hand,
when the source is an extended object approximately aligned with
the lens along the line of sight, the formation of arcs takes
place. The angular radius of the circle individuated by these arcs
is the Einstein angle $R_E$. Referring the interested reader to
the literature for more details, we here only remember that the
Einstein angle may be determined by solving\,:

\begin{equation}
\hat{\alpha}(\xi_E) = R_E = \xi_E R_{eff}, \label{eq: solxie}
\end{equation}
with\footnote{Here, we recast the definition for deflection angle
and projected mass introducing the adimensional radius $\xi_{E}$.}
$\xi_{E}=R_{E}/R_{eff}$, so that a measurement of $R_E$ (in $arcsec$)
provides a constraint on the model parameters through the deflection angle
$\hat{\alpha}(\xi_E)$. Moreover, comparing the definition of
$M_{proj}(\xi)$ and $\hat{\alpha}(\xi)$, it is immediate to get\,:

\begin{equation}
M_{proj}(R_E) = \pi R_E^2 \Sigma_{crit} = \pi \xi_E R_{eff}^2 \Sigma_{crit}
\label{eq: massprojobs}
\end{equation}
with $R_{eff}$ now expressed in linear units. Eq.(\ref{eq: massprojobs})
allows then to estimate the projected mass within the Einstein radius
provided that $\Sigma_{crit}$ is known, i.e., the lens and source redshift
have been measured.

\section{Testing the model global \ML\ ratio}\label{sec:MtoL_test}

Our proposed parametrization for $\Upsilon(r)$ was motivated by
the need to have a versatile expression able to mimic a large set
of galaxy models. Since we assume that the stellar component is
fixed and described by a PS density profile, mimicking a large set
of models is the same as mimicking the global \ML\ ratio predicted
by different dark halo models. Needless to say, such an ambitious
goal is difficult to reach so that we expect that Eq.(\ref{eq:
upsmodel}) does a good job over a limited radial range. It is
therefore worth exploring how large is this range and how good and
versatile is our approximated parametrization.

To this aim, it is convenient to resort to the following general
expression for the mass profile of the dark halo\,:
\begin{displaymath}
M_{DM}(r) = M_{vir} \times \mu(r/r_s) = M_{vir} \times \mu(\eta/\eta_s)
\end{displaymath}
with $M_{vir}$ the virial mass and $r_s = R_{eff} \eta_s $ a
scalelength. As a function of the adimensional length $\eta$ the
DM mass fraction then reads\,:

\begin{displaymath}
f_{DM}(\eta) =  \left [ 1 + \frac{M_{\star}(\eta)}{M_{DM}(\eta)}
\right ]^{-1}  \ .
\end{displaymath}
Normalizing with respect to $f_e = f_{DM}(R = R_{eff})$
and using Eq.(\ref{eq: lr}), we finally get\,:

\begin{eqnarray}
f_{DM}(\eta) & = & \left \{ 1 + \frac{1 - f_e}{f_e}
\frac{\mu(1/\eta_s)}{\mu(\eta)} \right . \nonumber \\
~ & \times & \left . \frac{1 + \Gamma[n(3 - p_n), b_n \eta^{1/n}]/\Gamma[n(3 - p_n)]}
{1 + \Gamma[n(3 - p_n),b_n]/\Gamma[n(3 - p_n)]} \right \}^{-1}
\label{eq: fdmgen}
\end{eqnarray}
which may then be related to the global \ML\ ratio as\,:

\begin{equation}
\Upsilon(\eta) = \frac{\Yst}{1 - f_{DM}(\eta)} \label{eq: upsfdm}
\end{equation}
having assumed a constant stellar \ML\ ratio. Eq.(\ref{eq:
fdmgen}) makes it possible to compute the DM mass fraction and
hence the global \ML\ ratio through Eq.(\ref{eq: upsfdm}) provided
that an expression for $\mu(\eta/\eta_s)$ is given.

To test the validity of our phenomenological ansatz, we start
considering the most popular galaxy profiles, i.e. NFW (Navarro et
al. 1996, 1997), the Einasto (\cite{E65,EH89,N03,PoLLS}), the
nonsingular isothermal sphere and the Burkert (1995) models. Using
the formalism introduced above, the cumulative mass profile for
each case is assigned by the following expressions\,:

\begin{displaymath}
\mu_{NFW}(\eta/\eta_s) = \log{(1 + \eta/\eta_s)} - (\eta/\eta_s)
(1 + \eta/\eta_s)^{-1},
\end{displaymath}

\begin{displaymath}
\mu_{Ein}(\eta) = 1 - \frac{\Gamma[2n_{DM}, d_n
(\eta/\eta_s)^{1/n_{DM}}]}{\Gamma(2n_{DM})} \ ,
\end{displaymath}

\begin{displaymath}
\mu_{NIS}(\eta) = \eta/\eta_s - \arctan{(\eta/\eta_s)} \ ,
\end{displaymath}

\begin{displaymath}
\mu_{Bur}(\eta) = \frac{\ln{[1 + (\eta/\eta_s)^2]}}{4} +
\frac{\ln{[1 + (\eta/\eta_s)]} - \arctan{(\eta/\eta_s)}}{2} \ .
\end{displaymath}
We parameterize NFW profile assigning first the virial mass
$M_{vir}$ and then estimating the scalelength $r_s$ as $r_s =
R_{vir}/c$ using the following numerically motivated relation
(\cite{bull01,Nap05}) to set the concentration parameter\,:

\begin{displaymath}
c = \frac{16.7}{1 + z} \left ( \frac{M_{vir}}{10^{11} h^{-1}
\ {\rm M_{\odot}}} \right )^{-0.125}
\end{displaymath}
also taking care of the large scatter. It is worth noting that, by
virtue of the $c$\,-\,$M_{vir}$ relation, the NFW profile reduces
to a one\,-\,parameter model. In contrast, since no
numerically or observationally motivated relation between
$(n_{DM}, \eta_s, M_{vir})$ is known, the Einasto model is
assigned by three parameters that we choose in such a way to have
a realistic density profile. A similar discussion also holds for
both the NIS and Burkert models which are, however,
two\,-\,parameters profiles. As a further remark, we stress that,
while the NIS and Burkert models possess an inner core, the NFW
and Einasto models are cuspy even if with different logarithmic
slope for $\eta << \eta_s$. Moreover, the outer slope of the
density profile runs from $s = -2$ for the NIS model, to $s = -3$
for both NFW and Burkert cases, to $s = -\infty$ (i.e. exponential
decay) for the Einasto parametrization. As such, by considering
these four models, we are spanning a wide range of possibilities
thus allowing us to test how good our phenomenological proposal
for the global \ML\ ratio $\Upsilon(r)$ mimics a large class of
different profiles.

Once the halo parameters have been set, we add a stellar component
fixing the photometric parameters $(n, R_{eff}, I_e)$, the stellar \ML\ ratio
\Yst\ and the redshift $z$ so that they are equal to those of a
randomly chosen lens. We then fit Eq.(\ref{eq: upsmodel}) to the
resulting $\Upsilon(\eta)$ profile and repeat this exercise for
1000 random realizations of each halo model. The instructive results of
this investigation are schematically summarized below.

\begin{itemize}

\item[-]{Eq.(\ref{eq: upsmodel}) fits the $\Upsilon(r)$ of the
tested models with a very good precision and over a large radial
range, namely $0.01 \le \eta \le 10$. Denoting with $\Upsilon_{fit}$ our
proposed function and defining $\Delta = | 1 - \Upsilon_{fit}/\Upsilon |$,
we get that the maximal deviation is lower than $15\%$ (depending on the model).
In particular, for NFW is $0.06\% \le \Delta_{rms} \le 1.8\%$ and
$2\% \le \Delta_{max} \le 8\%$. The Einasto model is fitted with
$\Delta_{rms} \sim 0.5\%$, while the worst performances are obtained
for the Burkert one for which we, however, find a still
satisfactory $\Delta_{rms} \sim 6\%$. \\}

\item[-]{The fitted model parameters $(\alpha, \beta, \eta_0, \Upsilon_{eff}/\Yst)$
depend on the halo parameters so that a general rule cannot be
given. However, we note that quite small and negative values of
$\alpha$ are clearly preferred for the NFW+PS input model. This is
somewhat surprising since this would give rise to models with an
inner asymptotic slope of the global density $s_0 = \alpha - 3 <
-3$, contrary to popular models having $s_0 \simeq -1$. However,
as stated above, in the very inner regions our model does not fit
anymore the NFW + PS one so that the results on $s_0$ can not be
trusted upon. For this same reason, we do not care about obtaining
still more negative values of $\alpha$. Indeed, for the other
models, $\alpha < 0$ cases are still clearly preferred. As a
conservative estimate, we therefore conclude that, in order our
parametrization fits well different dark halo models, one must
have negative $\alpha$, with values as low as $\alpha \simeq
-0.40$. \\}

\item[-]{The asymptotic outer slope $\alpha + \beta$ typically
takes values of order $1$\,-\,$2$, even if one cannot exclude
values as high as $\alpha + \beta \simeq 5$. For example, if
considering the NFW + PS, $\alpha + \beta$ is strongly peaked on
the $\sim 2$, and the distributions is markedly asymmetric with
long tails towards larger values. In contrast, $\alpha + \beta$
has a similar distribution, but shifted to the smaller $\alpha + \beta
\simeq 0.5$ peak, for cored models. As a general rule, however, $\alpha +
\beta \le 0$ never occurs in agreement with our above discussion
showing that such models have an unphysically decreasing mass
profile. \\}

\item[-]{The scaling quantities $\Upsilon_{eff}/\Yst$ and $\log{\eta_0}$
are in the range $1.0 \le \Upsilon_{eff}/\Yst \le 2.5$ and $-0.5 \le
\log{\eta_0} \le 1.5$. In particular, if considering the NFW + PS,
$\Upsilon_{eff}/\Yst$ and $\log{\eta_0}$ are strongly peaked on
$\sim 1.5$ and $0.5$ respectively. Similar values are obtained for the
other dark halo models considered, a result which is not fully unexpected.
Indeed, $\Upsilon_{eff}/\Upsilon_{\star}$ and $\log{\eta_0}$ are mostly
related respectively to the virial mass $M_{vir}$ and DM mass fration $f_e$
and to the scaled virial radii $R_{vir}/R_{eff}$. Since these quantities take
similar values for all models, we indeed expect to find a not too large scatter
among different models which is indeed what we find. \\}

\item[-]{Changing the $c$\,-\,$M_{vir}$ relation or its scatter does
not alter the result on the validity of our approximation for NFW + PS profile,
i.e. $\Delta_{rms}$ and $\Delta_{max}$ span the same range as above for
$\eta$ covering the same radial range. However, the distribution
of the fitted parameters may change significantly. In particular,
higher values of $\alpha$ may be obtained although negative
$\alpha$ are still preferred\footnote{Note that the relation between the
concentration $c$ and $\alpha$ is difficult to extract. Indeed, since we
set $\alpha$ as a result of fitting our phenomenological model to an input
stellar plus dark halo profile, its value also depends on the fitted
values of the other parameters, $(\alpha + \beta, \eta_0)$. As a result,
we are therefore unable to say whether a larger $\alpha$ implies a higher
or smaller concentration.}. Considering that similar values for $\alpha$
can be obtained by fitting $\Upsilon_{fit}$ to models other than NFW, we
argue that it is not possible to resort to $\alpha$ to discriminate between,
e.g., the NFW and the Einasto models, or cusped and cored profiles, unless
one has a precise determination of the $c$\,-\,$M_{vir}$ relation. \\}

\end{itemize}
These results clearly show that our proposed parametrization of
the global \ML\ ratio indeed makes very well its job. It is indeed
able to reproduce with very good accuracy the expected behaviour
of $M(r)/L(r)$ whatever is the dark halo density profile over a
radial range spanning three orders of magnitude. As a final
cautionary test, we check that using this very good approximation
does not introduce any bias in the lensing deflection angle. As an
example, we can consider the case of the NFW model and comparing
the value of $\hat{\alpha}(\xi)$ computed using the analytical NFW
deflection angle and the numerically evaluated Eq.(\ref{eq:
defangle}). We repeat this test for 1000 random realizations and
estimate the ratio
$\hat{\alpha}_{NFW}(\xi)/\hat{\alpha}_{fit}(\xi)$ for $\xi = 1$
and $\xi = \xi_E$, using for $\xi_E$ the observational value for
the lens used to set the stellar parameters. It turns out that
both these ratios deviate from unity less than $1\%$ confirming
our expectation that no bias is induced by the use of our ansatz.

\section{Model vs observations}\label{sec:mod_vs_obs}

The phenomenological ansatz for $\Upsilon(r)$ we are proposing is
characterized by four parameters, namely the two asymptotic
slopes\footnote{Hereafter, we will consider $\alpha + \beta$
rather than $\beta$ as model parameter since it is this former
quantity that sets the asymptotic slope of the global \ML\ ratio.
Moreover, we have physically motivated constraints on $\alpha$ and
$\alpha + \beta$ so that the ranges to be explored for these
parameters is immediately set. We also change from $\eta_0$ to
$\log{\eta_0}$ as scalength parameter in order to explore a wider
range for this quantity.}, the logarithm of the scaling length
$\log{\eta_0}$ and the global \ML\ ratio at the effective radius
$\Upsilon_{eff}$. To these four quantities, we have to add the
stellar \ML\ ratio \Yst\ thus leading to five the number of
unknown quantities to be constrained. Needless to say, tackling
such an issue is a quite daunting task. A possible way out could
be contrasting the model with kinematic observations, such as the
velocity dispersion profile. To this aim, the data should cover a
radial range wide enough to sample with sufficient detail both the
inner and the outer regions in order to constrain the two
asymptotic slopes. Moreover, the measurement errors and the
sampling should be extremely good to overcome the problem of
parameters degeneracies in a five dimensional space. All these
observational requirements are quite demanding so as to be
satisfied only by a handful number of nearby galaxies. Actually,
we are interested here in lens galaxies with typical redshifts of
order $0.1 - 0.5$ so that measuring their velocity dispersion
profile with the above precision and sampling is an unrealistic
task. Nevertheless, for each galaxy, we still have two observable
quantities that can be used, namely the Einstein angle $R_E$ and
$\sigma_{ap}$, the line of sight velocity dispersion luminosity
weighted in a circular aperture of radius $R_{ap}$. Needless to
say, it is impossible to constraint a five dimensional parameter
space with only two datapoints so that a case\,-\,by\,-\,case
analysis is not possible. In order to overcome such a problem, we
have therefore to reduce the number of unknowns and increase the
number of observed datapoints. To this aim, we therefore first get
an estimate of the stellar \ML\ ratio \Yst.

\subsection{Stellar M/L}

We start assembling a library of synthetic stellar population
models obtained through the {\tt Galaxev} code (\cite{BC03})
varying the age of the population, its metallicity and time lag of
the exponential star formation rate and assuming a Chabrier (2001)
initial mass function (IMF). Then, we use the tabulated $(u, g, r,
i, z)$ apparent magnitudes (corrected for extinction) of each lens
to fit the above library of spectra (suitably redshifted to lens
redshift) to the colours, thus getting the estimates reported in
Table 1. Note that these values may be easily scaled to a Salpeter
(1955) or Kroupa (2001) IMF by multiplying by $1.8$ or $1.125$
respectively so that we can explore other IMF
choices\footnote{While this is correct for a Salpeter IMF, since
it differs from a Chabrier IMF only for the low mass slope and
predict very similar colours, for a Kroupa IMF this is not
strictly true, but we could assume the scale factor above as a
good approximation.}. In order to get the \Yst\ uncertainties, we
use a Monte Carlo\,-\,like procedure generating a set of colours
from a Gaussian distribution centred on each mean colour and
standard deviation equal to the colour uncertainty. Fitting the
colours thus obtained to the synthetic spectra for each
realization, we generate a distribution of fitted parameters. The
median and median scatter of such a distribution is finally taken
as an estimate of \Yst\ and its uncertainty (see Tortora et al.
2009 for further details).

\subsection{Resorting to universal parameters}

Having reduced by one the number of parameters, we now look for a
way to increase the number of constraints. To this aim, we could
stack all the lenses in a single sample. However, while one can
reasonably argue that the same functional expression for
$\Upsilon(r)$ describes the global \ML\ ratio of all the lenses,
its characterizing parameters are likely to change on a
case\,-\,by\,-\,case basis. Indeed, even if we assume that the DM
halo has a universal profile, the details of the baryonic assembly
may lead to different model parameters. In order to partially
alleviate this problem, we therefore reparametrize $\Upsilon(r)$
in Eq.(\ref{eq: upsmodel}) in terms of quantities that are likely
to be (at least, to first order) universal.

To this aim, it is worth remembering that previous works on
fitting galaxy models to the lenses using the constraints from the
Einstein ring and the velocity dispersion concordantly suggest
that the total mass profile at $R = R_{E}$ is well approximated by
a singular isothermal sphere (see, e.g., \cite{TK04} and refs.
therein). Actually, rather than constraining the global mass
profile, such an analysis essentially probe the shape of the
density profile only at the Einstein radius so that one can argue
that the logarithmic slope of  the density profile at $R_E/R_{eff}$ is the
same for all lenses. Motivated by these previous literature
results, we therefore assume that $s_E = s(R_E/R_{eff})$ is the
same for all lenses and use this quantity as model parameter
instead of $\log{\eta_0}$. To this aim, for given values of
$(\alpha, \alpha + \beta, s_E)$, we numerically solve\,:

\begin{displaymath}
s(R_E/R_{eff}) = s_E
\end{displaymath}
with respect to $\log{\eta_0}$ using  Eqs.(\ref{eq:
seta})\,-\,(\ref{eq: d2}) to compute $s(\eta)$. We then note that
the DM mass fraction at the virial radius is likely to be a
universal quantity. Indeed, considering that\,:

\begin{displaymath}
f_{vir} = 1 - \frac{M_{\star}(R_{vir})}{M_{vir}} \simeq 1 -
\frac{\Yst\ L_T}{M_{vir}} \ ,
\end{displaymath}
and setting $M_{\star}(R_{vir})/M_{vir} = \varepsilon_{SF} f_{b}$
with $f_{b}=\Omega_b/\Omega_M$ the cosmic baryon fraction
(\cite{WMAP}) and $\varepsilon_{SF}$ the star formation
efficiency, we can assume that, notwithstanding the details of the
star formation process, $\varepsilon_{SF}$ is roughly the same for
all the lenses in the sample. Although not strictly true, this is,
however, a resonable approximation since all the lenses we are
considering are elliptical galaxies spanning a limited range in
both redshift and luminosity. As such, we will hereafter assume
that $\mu_{vir} = M_{vir}/\Yst L_T$ is a universal quantity to set
the $\Upsilon_{eff}$ parameter as\,:

\begin{equation}
\Upsilon_{eff} = \frac{M_{vir}}{\eta_0^{\alpha} L_T}
\left ( \frac{1 + \eta_0}{\eta_{vir} + \eta_0} \right )^{\beta}
\frac{\Gamma[n(3 - p_n)]}{\gamma[n(3 - p_n), b_n \eta_{vir}^{1/n}]}
\label{eq: upseffmvir}
\end{equation}
with $M_{vir} = \mu_{vir} \Yst L_T$. Note that, here, we are
assuming that $M_{\star}(R_{vir})$ is the total stellar mass $\Yst
L_T$ which, although not strictly true, is a very good
approximation.

Unfortunately, we are unable to find other lens related parameters
that can be considered universal so that we will hereafter
parameterize our $\Upsilon(r)$ ansatz with the slope parameters
$\alpha$ and $\alpha + \beta$, the logarithmic slope at the scaled
Eintein radius $s_E = s(R_E/R_{eff})$, and the mass ratio at the
virial radius $\mu_{vir} = M_{vir}/\Yst L_T$ assuming that these
are universal quantities. In order to explore the impact of this
theoretical hypothesis, we will fit our model to the full lens
sample and to four subsamples containing respectively five, six,
six, four lenses binned according to the absolute $V$ magnitude
$M_V$. Comparing the constraints on the model parameters obtained
by the different fits makes it possible to look for an eventual
dependence on the lens luminosity thus giving an a posteriori
check of our a priori assumption.

\subsection{Statistical analysis}

As shown by Eq.(\ref{eq: solxie}), a measurement of $\xi_E$ may be
considered as a measurement of the deflection angle $\hat{\alpha}(\xi_E)$
so that stacking together many galaxies with different values of $\xi_E$
makes it possible to reconstruct the deflection angle profile. Similarly, since
$\sigma_{ap}$ is weighted within an aperture of fixed angular
radius $R_{ap}$ but different normalized radii $\xi_{ap} =
R_{ap}/R_{eff}$, this is the same as tracing the luminosity weighted line of
sight dispersion profile. The agreement within the model and the
data may then be optimized by maximizing the likelihood
function\,:

\begin{equation}
{\cal{L}}({\bf p}) \propto \exp{\left [- \frac{\chi^2_{lens}({\bf p}) +
\chi^2_{dyn}({\bf p})}{2} \right ]}
\label{eq: deflike}
\end{equation}
with ${\bf p} = (\alpha, \alpha + \beta, s_E, \mu_{vir})$ the set
of model parameters and $\chi^2_{lens}$ ($\chi^2_{dyn}$) a merit
function for the lensing (dynamics) related observables.
Maximizing ${\cal{L}}({\bf p})$ gives us the best fit model
parameters, while the isolikelihood contours provide constraints
on the parameter space that are difficult to quantify because of
the four dimensional nature of this space. In order to carry out
the maximization of ${\cal{L}}({\bf p})$, we implement a Markov
Chain Monte Carlo algorithm which efficiently samples the regions
of high likelihood. The points in the chain may be also used to
extract constraints on a given parameter $p_i$ marginalizing over
the other ones. To this aim, we consider the histogram of the
$p_i$ values along the chain and, following a Bayesian approach,
take the median of this histogram as the best estimate of the
parameter $p_i$, while $68$ and $95\%$ quantiles of this
distribution give us the $68$ and $95\%$ confidence ranges for the
parameter $p_i$. Note that, since the distributions may also be
not Gaussian, the median and the mean value may be different, but
we still retain the median as best estimator in agreement with the
Bayesian framework. As a further remark, let us stress that the
best fit parameters ${\bf p}_{BF}$ may also significantly differ
from the maximum likelihood values ${\bf p}_{ML}$ because of
parameter degeneracies. Actually, in order to have ${\bf p}_{BF} =
{\bf p}_{ML}$, ${\cal{L}}(\bf p)$ should be written as the product
of the marginalized likelihood functions. Moreover, these latter
should be Gaussian so that they are centred on the maximum
likelihood values (therefore coincident with the median values).
Actually, both these conditions are rarely met so that having
${\bf p}_{BF} \ne {\bf p}_{ML}$ is quite common.

The Markov Chains may also be used to extract constraints on some
interesting derived quantities. To this aim, let us consider a
generic function $y = g({\bf p})$. Evaluating $y$ along the
chains, we can build up the histogram of its values and then use
the median and quantiles of this latter to infer the Bayesian
confidence levels on $y$. Such a procedure will be used for the
scaling model parameters $\log{\eta_0}$ and $\Upsilon_{eff}$ and
for the projected and spherical DM mass fractions for each lens in
the sample.

\begin{table*}
\centering
\begin{tabular}{cccccccccc}
\hline Name & $z_L$ & $z_S$ & $M_V$ & $R_{eff}$ & $R_E$ &
$\sigma_{ap}$ & $R_{ap}/R_{eff}$ & \Yst\ \\
\hline \hline

$~$ & $~$ & $~$ & $~$ & $~$ & $~$ & $~$ & $~$ & $~$ \\

J002907.8-005550 & 0.2270 & 0.9313 & -21.53 & 1.48 & 0.82 &
$229 \ {\pm} \ 18$ & $1.01$ & $1.85 \ {\pm} \ 0.15$ \\

J015758.9-005626 & 0.5132 & 0.9243 & -22.16 & 0.93 & 0.72 &
$295 \ {\pm} \ 47$ & $1.61$ & $2.00 \ {\pm} \ 0.45$ \\

J021652.5-081345 & 0.3317 & 0.5235 & -22.95 & 2.79 & 1.15 &
$333 \ {\pm} \ 23$ & $0.54$ & $2.12 \ {\pm} \ 0.09$ \\

J025245.2+00358 & 0.2803 & 0.9818 & -21.67 & 1.69 & 0.98 &
$164 \ {\pm} \ 12$ & $0.89$ & $2.19 \ {\pm} \ 0.29$ \\

J033012.1-002052 & 0.3507 & 1.0709 & -21.86 & 1.17 & 1.06 &
$212 \ {\pm} \ 21$ & $1.28$ & $2.49 \ {\pm} \ 0.15$ \\

J072805.0+383526 & 0.2058 & 0.6877 & -21.80 & 1.33 & 1.25 &
$214 \ {\pm} \ 11$ & $1.13$ & $1.80 \ {\pm} \ 0.15$ \\

J080858.8+470639 & 0.2195 & 1.0251 & -21.77 & 1.65 & 1.23 &
$236 \ {\pm} \ 11$ & $0.91$ & $2.20 \ {\pm} \ 0.15$ \\

J090315.2+411609 & 0.4304 & 1.0645 & -22.25 & 1.28 & 1.13 &
$223 \ {\pm} \ 27$ & $1.17$ & $2.09 \ {\pm} \ 0.15$ \\

J091205.3+002901 & 0.1642 & 0.3239 & -22.95 & 5.50 & 1.61 &
$326 \ {\pm} \ 16$ & $0.27$ & $1.45 \ {\pm} \ 0.07$ \\

J095944.1+041017 & 0.1260 & 0.5350 & -20.94 & 1.99 & 1.00 &
$197 \ {\pm} \ 13$ & $0.75$ & $2.09 \ {\pm} \ 0.10$ \\

J102332.3+423002 & 0.1912 & 0.6960 & -21.56 & 1.40 & 1.30 &
$242 \ {\pm} \ 15$ & $1.07$ & $2.54 \ {\pm} \ 0.09$ \\

J110308.2+532228 & 0.1582 & 0.7353 & -22.02 & 3.22 & 0.84 &
$196 \ {\pm} \ 12$ & $0.46$ & $1.10 \ {\pm} \ 0.06$ \\

J120540.4+491029 & 0.2150 & 0.4808 & -22.00 & 1.92 & 1.04 &
$281 \ {\pm} \ 14$ & $0.78$ & $1.98 \ {\pm} \ 0.16$ \\

J125028.3+052349 & 0.2318 & 0.7953 & -22.17 & 1.64 & 1.15 &
$252 \ {\pm} \ 14$ & $0.91$ & $2.18 \ {\pm} \ 0.10$ \\

J140228.1+632133 & 0.2046 & 0.4814 & -22.20 & 2.29 & 1.39 &
$267 \ {\pm} \ 17$ & $0.65$ & $1.50 \ {\pm} \ 0.08$ \\

J142015.9+601915 & 0.0629 & 0.5351 & -21.04 & 2.49 & 1.04 &
$205 \ {\pm} \ 10$ & $0.60$ & $1.24 \ {\pm} \ 0.10$ \\

J162746.5-005358 & 0.2076 & 0.5241 & -22.06 & 2.47 & 1.21 &
$290 \ {\pm} \ 15$ & $0.61$ & $1.80 \ {\pm} \ 0.89$ \\

J163028.2+452036 & 0.2479 & 0.7933 & -22.31 & 2.01 & 1.81 &
$276 \ {\pm} \ 16$ & $0.75$ & $2.42 \ {\pm} \ 0.09$ \\


J230053.2+002238 & 0.2285 & 0.4635 & -22.06 & 2.22 & 1.25 &
$279 \ {\pm} \ 17$ & $0.68$ & $1.98 \ {\pm} \ 0.16$ \\

J230321.7+142218 & 0.1553 & 0.5170 & -22.40 & 3.73 & 1.64
& $255 \ {\pm} \ 16$ & $0.40$ & $2.49 \ {\pm} \ 0.08$ \\

J234111.6+000019 & 0.1860 & 0.8070 & -22.14 & 3.20 & 1.28 &
$207 \ {\pm} \ 13$ & $0.47$ & $2.89 \ {\pm} \ 0.15$ \\

$~$ & $~$ & $~$ & $~$ & $~$ & $~$ & $~$ & $~$ & $~$ \\
\hline
\end{tabular}
\caption{Photometric and lensing observables and estimated stellar
\ML\ ratio for the 21 SLACS lenses.}
\end{table*}

\subsubsection{The lensing merit function}

Dealing with the Einstein angle as a constraint is not
straightforward. Indeed, for a given set of model parameters,
$\xi_E$ should be find solving Eq.(\ref{eq: solxie}) so that a lot
of computing time should be spent to explore the four dimensional
parameter space. Moreover, the photometric parameters $(n, I_e,
R_{eff})$ entering Eq.(\ref{eq: solxie}) are affected by their own
uncertainties so that one should solve this relation many times
by, for instance, bootstrapping the values of these quantities and
looking for the corresponding uncertainty induced on the predicted
$\xi_E$. Fortunately, we can skip all these complications by
resorting to the projected mass within the Einstein radius whose
theoretical and observed values may be estimated from
Eqs.(\ref{eq: massprojth}) and (\ref{eq: massprojobs}).

Because of the measurement errors on $I_e$ and $R_{eff}$ and the
uncertainty on $\Upsilon_{eff}$ coming from the one on \Yst, the
predicted $M_{proj}^{th}(\xi_E)$ is affected by an uncertainty
which we naively quantity using error propagation as\,:

\begin{equation}
\frac{\delta M_{proj}^{th}(\xi_E)}{M_{proj}^{th}(\xi_E)} =
\sqrt{\left ( \frac{\delta L_T}{L_T} \right )^2 + \left (
\frac{\delta \Upsilon_{eff}}{\Upsilon_{eff}} \right )^2}
\label{eq: mptherr}
\end{equation}
with\,:

\begin{displaymath}
\frac{\delta L_T}{L_T} = \ln{10} \sqrt{(2 \delta \log{R_{eff}})^2
+ (\delta \log{I_e})^2} \ ,
\end{displaymath}

\begin{displaymath}
\frac{\delta \Upsilon_{eff}}{\Upsilon_{eff}} = \sqrt{\left (
\frac{\delta L_T}{L_T} \right )^2 + \left ( \frac{\delta
M_{vir}}{M_{vir}} \right )^2} \ ,
\end{displaymath}

\begin{displaymath}
\frac{\delta M_{vir}}{M_{vir}} = \sqrt{\left ( \frac{\delta
\Yst}{\Yst} \right )^2 + \left ( \frac{\delta L_T}{L_T} \right
)^2} \ ,
\end{displaymath}
Note that we have here neglected the contribution of the error on
$R_E$ entering through the integral in Eq.(\ref{eq: massprojth}).
To take this into account, we should compute the numerical
derivative $dM_{proj}^{th}(\xi)/d\xi$ for each choice of the model
parameters. Such a computationally expensive procedure may be
however skipped since, being $R_E$ measured with a quite low
error, the integral function may be well approximated as a
constant over such a narrow range. As a consequence, the
corresponding error term is safely negligible with respect to the
ones due to the measurement uncertainties on $I_e$ and $R_{eff}$.
We stress, however, that neglecting this term makes our analysis
more conservative since we are actually assuming that the final
uncertainty on $M_{proj}^{th}(\xi_E)$ is lower than its true
value. The error on $R_E$ is instead taken into account to compute
the one on the observed projected mass being\,:

\begin{equation}
\delta M_{proj}^{obs}(\xi_E)/M_{proj}^{obs}(\xi_E) = 2 \delta
R_E/R_E \ . \label{eq: mpobserr}
\end{equation}
We finally define the lensing merit function as\,:

\begin{equation}
\chi^2_{lens}({\bf p}) = \sum_{i =  1}^{{\cal{N}}}{\left [
\frac{M_{proj,i}^{obs}(\xi_E) - M_{proj,i}^{th}(\xi_E, {\bf p})}
{\varepsilon_{lens,i}} \right ]^2} \label{eq: defchilens}
\end{equation}
where the total uncertainty $\varepsilon_{lens,i}$ is given by the
sum in quadrature of the theoretical and observational
uncertainties estimated from Eqs.(\ref{eq: mptherr}) and (\ref{eq:
mpobserr}) and the sum is over the ${\cal{N}}$ lenses in the
sample.

\subsubsection{The dynamics figure of merit}

Denoting with $\sigma_{ap,obs}$ the observed value, we define the
following figure of merit for the dynamical data\,:

\begin{equation}
\chi^2_{dyn}({\bf p}) = \sum_{i = 1}^{{\cal{N}}} {\left [
\frac{\sigma_{ap,obs}^{(i)} - \sigma_{ap,th}^{(i)}({\bf p})}
{\varepsilon_{dyn,i}} \right ]^2} \label{eq: defchidyn}
\end{equation}
where $\varepsilon_{dyn}^{(i)}$ is obtained by summing in
quadrature the theoretical and measurement uncertainties, while
the sum is, as usual, over the ${\cal{N}}$ objects in the sample.
Note that, although $R_{ap}$ is the same for all galaxies, the
different measurements of $\sigma_{ap}$ actually probe a range of
scaled radii $\xi_{ap}$ since the effective radius changes on a
case\,-\,by\,-\,case basis. As a final remark, it is worth
stressing that $\sigma_{ap,th} \propto \Upsilon_{eff}^{1/2}$,
while $M_{proj}^{th}(\xi_E) \propto \Upsilon_{eff}$. It is this
diverse scaling with $\Upsilon_{eff}$ (and the other model
parameters too) which makes combining lensing and dynamical data
an efficient tool to partially break the degeneracies among model
parameters thus strengthening the constraints we can obtain.

\subsection{The data}

The sums in Eqs.(\ref{eq: defchilens}) and (\ref{eq: defchidyn})
run over the ${\cal{N}}$ objects in the sample. It is therefore
vital to specify what this sample is made of. We consider here the
21 lens ETGs reported in Gavazzi et al. (2007) whose main properties
are summarized in Table 1 where columns are as follows\,: 1.
name (without the prefix SDSS); 2. lens redshift; 3. source
redshift; 4. absolute $V$ magnitude (incremented by $5 \log{h}$);
5. effective radius (in $arcsec$); 6. Einstein radius (in
$arcsec$); 7. aperture velocity dispersion (in km/s); 8. ratio
between the aperture and the effective radii; 9. estimated V-band
stellar \ML\ ratio with its error. Note that we do not have colours
for lens J223840.2+075456 so that \Yst\ cannot be evaluated and
this object will be excluded by the analysis. These lenses have been
observed in the framework of the SLACS survey which aims at
confirming through ACS imaging candidate lenses spectroscopically
identified within the SDSS catalog (see Bolton et al. 2004, 2006,
2008 for further details). The SDSS velocity dispersions have been
measured within a circular aperture of fixed radius $R_{ap} = 1.5
\, \rm arcsec$ and spans the range $196 \le \sigma_{ap} \le 333 \
{\rm km/s}$ with a mean square velocity $\langle \sigma_{ap}^2
\rangle^{1/2} \simeq 248 \ {\rm km/s}$. The lens galaxies have a
mean redshift $\langle z_L \rangle \simeq 0.22$, but it is worth
noting that the redshift range covered ($0.063 \le z_L \le 0.513$)
probes almost one order of magnitude (even if quite sparsely). The
measured surface brightness has been fitted by a de Vaucouleurs
(1948) model so that $n$ will be fixed to 4 in the following.
Typical uncertainties on $R_{eff}$ and $I_e$ are quite small so
that, following Bolton et al. (2008), we will set $(\delta
\log{I_e}, \delta \log{R_{eff}}) = (0.020, 0.015)$ for all
galaxies in the sample. Gavazzi et al. (2007) provides also
absolute $V$ band magnitude corrected for filter transformation
and Galactic extinction and $K$ and evolution corrected to a
common redshift $z = 0.2$. When fitting Einstein radius and
velocity dispersions, however, we need the luminosity at the lens
redshift. This can be easily estimated as (\cite{Gav07})\,:

\begin{displaymath}
\log{L_V}(z_L) = \log{L_V}(z = 0.2) - 0.40 (z_L - 0.2)
\end{displaymath}
with\,:

\begin{displaymath}
\log{L_V}(z = 0.2) = {\rm dex}\left ( -\frac{M_V - M_{V,\odot}}{2.5} \right )
\end{displaymath}
having set ${\rm dex}(x) = 10^x$ and denoted by $M_V$ and $M_{V,\odot}
= 4.83$ the absolute magnitudes of the lens at $z = 0.2$ and of the Sun, respectively.
We further estimate the effective intensity as (\cite{GD05})\,:

\begin{equation}
I_e = \frac{\langle I_e \rangle}{n {\rm e}^{b_n} b_n^{-2n}
\Gamma(2n)}
\label{eq: ieest}
\end{equation}
with the effective surface brightness given by\,:

\begin{equation}
\langle I_e \rangle = L_V(z_L)/2 \pi R_{eff}^2 \ .
\label{eq: defiemean}
\end{equation}
While $\sigma_{ap}$ is directly measured from the lens spectrum,
the Einstein radii are determined by a parameterized procedure.
The lens potential is modeled as a singular isothermal ellipse,
while a non parametric technique is employed to reconstruct the
source surface brightness profile. The Einstein radii is then
estimated so that the lensed source matches the observed
arc\,-\,like features. Typical uncertainties in $R_E$ are of order
$0.05 \, \rm arcsec$ with a small variation among the lenses.
Considering the different $R_E$ values, we naively set $\delta
R_E/R_E = 0.05$ for all the sample objects. Table 1 summarizes the
relevant quantities for our analysis.

\section{Results}\label{sec:results}

The sample of 21 SLACS lenses represents the input dataset needed to
constrain the four model parameters $(\alpha, \alpha + \beta,
s_E, \mu_{vir})$ through the Bayesian likelihood analysis described above.
As a first test, motivated by the above discussion, we fit the model to the
full sample without any binning in luminosity thus giving us a quite
large set of constraints (namely, $2 {\times} 21 = 42$ observable quantities vs 4
parameters). Nevertheless, some care is needed when examining the results
of the Markov Chain analysis.

In order the results to be reliable, one should carefully check
that the chains have reached convergence, that is to say that the
chains have fully explored the regions of high likelihood. Should
this be the case, one should see the points of the chains for each
single parameter oscillate around an average value or, put another
way, the histograms of the parameters be single peaked
(eventually, with a Gaussian shape). In order to test the
convergence of the chains we resort to the test described in
Dunkley et al. (2005, see also \cite{Du09}) which is based on the
analysis of the chain power spectra. Before checking for
convegence, however, we first cut out the initial $30\%$ of the
chain in order to avoid the burn in period. Moreover, to reduce
spurious correlations among parameters induced by the nature of
the Markov Chain process, we thin the chain by taking 1 point each
25. We find that the convergence test is well passed for a chain
containing $10^{5}$ points which reduces to 2801 after the burn in
cut and the thinning. Such a large sample is more than sufficient
to sample the four dimensional parameter space. Note that we have
also tried to change the burn in cut and thinning thus checking
that the results are unaffected by these (somewhat arbitrary)
choices.

\subsection{Constraints on the model parameters}

The results obtained fitting our model to the full SLACS sample are
summarized in Table 2 which reports, for each parameter, mean and median 
values and $68$ and $95\%$ confidence limits. First, we note that the 
best fit model parameters turns out to be\,:

\begin{table}
\begin{center}
\begin{tabular}{ccccc}
\hline
Par & mean & median & $68\%$ CL & $95\%$ CL \\
\hline \hline
$~$ & $~$ & $~$ & $~$ & $~$ \\

$\alpha$ & $-0.26$ & $-0.25$ & $(-0.36, -0.16)$ & $(-0.43, -0.06)$ \\

$~$ & $~$ & $~$ & $~$ & $~$ \\

$\alpha + \beta$ & $0.55$ & $0.54$ & $(0.47, 0.63)$ & $(0.40, 0.71)$ \\

$~$ & $~$ & $~$ & $~$ & $~$ \\

$s_E$ & $-2.17$ & $-2.17$ & $(-2.24, -2.10)$ & $(-2.31, -2.03)$ \\

$~$ & $~$ & $~$ & $~$ & $~$ \\

$\mu_{vir}$ & $13.60$ & $12.96$ & $(9.52, 17.81)$ & $(7.60, 23.29)$ \\

$~$ & $~$ & $~$ & $~$ & $~$ \\
\hline
\end{tabular}
\end{center}
\caption{Results for the fit to the full lens sample.}
\end{table}

\begin{displaymath}
(\alpha, \alpha + \beta, s_E, \mu_{vir}) =
(-0.34, 0.52, -2.16, 12.22)
\end{displaymath}
giving $\chi^2/d.o.f. = 63.34/38 = 1.66$. From a statistical point of
view, such a reduced $\chi^2$ value could seem too large. Actually, a large
$\chi^2/d.o.f.$ may be due to a model failure or to an intrinsic scatter
which has not been taken into account. Such a second possibility is indeed
the more realistic considering that, having stacked all the lens together
notwithstanding their different photometric and mass properties, a certain
degree of scatter is indeed expected. Moreover, the overall quality of the
fit appears to be quite good. In order to quantify the agreement between
the data and the model for the best fit parameters, we first introduce
the two following quantities\,:

\begin{displaymath}
\frac{\Delta M_E}{\varepsilon_{lens}} = \frac{M_{proj}^{obs}(\xi_E)
- M_{proj}^{th}(\xi_E, {\bf p}_{bf})}{\varepsilon_{lens}} \ ,
\end{displaymath}

\begin{displaymath}
\frac{\Delta \sigma_{ap}}{\varepsilon_{dyn}} =
\frac{\sigma_{ap,obs} - \sigma_{ap,th}({\bf p}_{bf})}
{\varepsilon_{dyn}}
\end{displaymath}
with ${\bf p}_{bf}$ the set of best fit parameters. It turns out that\,:

\begin{displaymath}
\langle \Delta M_E/\varepsilon_{lens} \rangle = -0.12 \ \ ,
\ \ rms(\Delta M_E/\varepsilon_E) = 1.09 \ \ ,
\end{displaymath}

\begin{displaymath}
\langle \Delta \sigma_{ap}/\varepsilon_{dyn} \rangle = -0.04 \ \ , \
\ rms(\Delta \sigma_{ap}/\varepsilon_{dyn}) = 1.35 \ \ .
\end{displaymath}

\begin{table*}
\begin{center}
\begin{tabular}{cccccc}
\hline
Sample & $M_V$ & $(\alpha, \alpha + \beta, s_E, \mu_{vir})_{bf}$ & $\Delta M_E/\varepsilon_{lens}$
& $\Delta \sigma_{ap}/\varepsilon_{dyn}$ & $\chi^2/d.o.f.$ \\
\hline \hline

$~$ & $~$ & $~$ & $~$ & $~$ & $~$ \\

B1 & $-22.40_{-0.55}^{+0.15}$ & (-0.12, 1.31, -2.19, 72.42)
& (-0.12, 0.58) & (-0.11, 0.57) & 0.55  \\

$~$ & $~$ & $~$ & $~$ & $~$ & $~$ \\

B2 & $-22.15_{-0.05}^{+0.09}$ & (-0.45, 1.22, -2.20, 73.30)
& (-0.16, 1.42) & (-0.06, 1.45) & 3.08 \\

$~$ & $~$ & $~$ & $~$ & $~$ & $~$ \\

B3 & $-21.83_{-0.19}^{+0.16}$ & (-0.41, 0.44, -2.20, 8.51)
& (-0.03, 0.55) & (-0.12, 1.11) & 1.15 \\

$~$ & $~$ & $~$ & $~$ & $~$ & $~$ \\

B4 & $-21.33_{-0.25}^{+0.39}$ & (-0.44, 0.33, -2.39, 5.00)
& (-0.11, 0.65) & (-0.04, 0.34) & 0.55 \\

$~$ & $~$ & $~$ & $~$ & $~$ & $~$ \\

\hline
\end{tabular}
\end{center}
\caption{Results for the fit to the binned lenses.}
\end{table*}

\noindent The low values of $\langle \Delta M_E/\varepsilon_{lens}
\rangle$ and $\langle \Delta \sigma_{ap}/\varepsilon_{dyn}
\rangle$ are a clear evidence that the model is working well,
while the larger rms values are expected considering that we are
stacking galaxies with different intrinsic and environmental
properties. It is worth stressing, moreover, that both the
projected mass within the Einstein ring and the aperture velocity
dispersion predicted by the best fit model is well within $1
\sigma$ from the observed values for most of the galaxies, while
there is only one lens\footnote{This is the lens identified as
SDSS J234111.6+000019. In all the fits we will consider, this is,
indeed, the lens showing the larger normalized residuals for both
the projected mass and the velocity dispersion so that it is
possible that some peculiar feature is at work. However, in order
to not reduce the number of constraints without a definite
motivation, we prefer to retain this object in the sample even if
we caution the reader againt overrating its large residuals.} with
a theoretically predicted velocity dispersion $3 \sigma$ larger
than the observed value. We therefore safely conclude that the
model have successfully fitted the data for the lens sample we are
considering.

\begin{figure*}
\centering
\subfigure{\includegraphics[width=8cm]{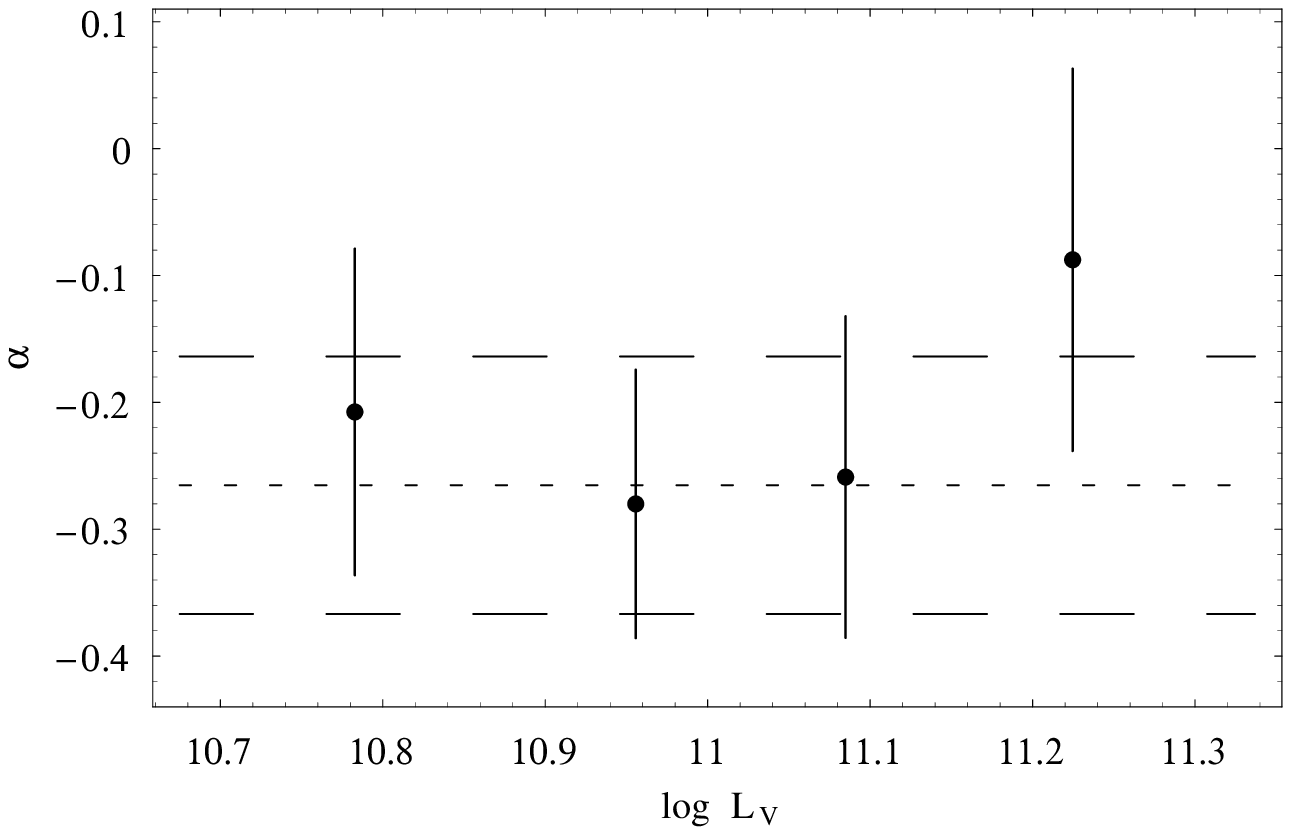}}
\goodgap \subfigure{\includegraphics[width=8cm]{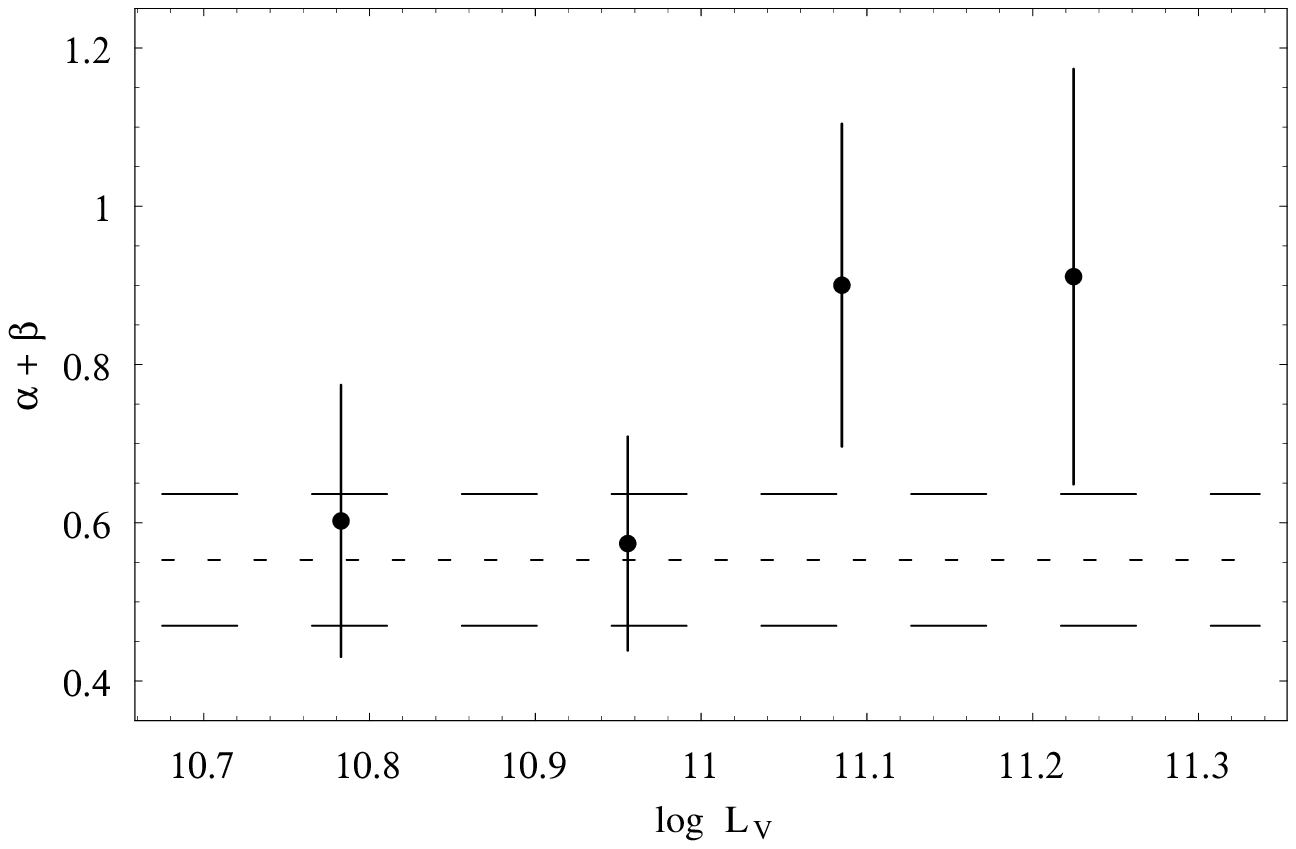}} \\
\subfigure{\includegraphics[width=8cm]{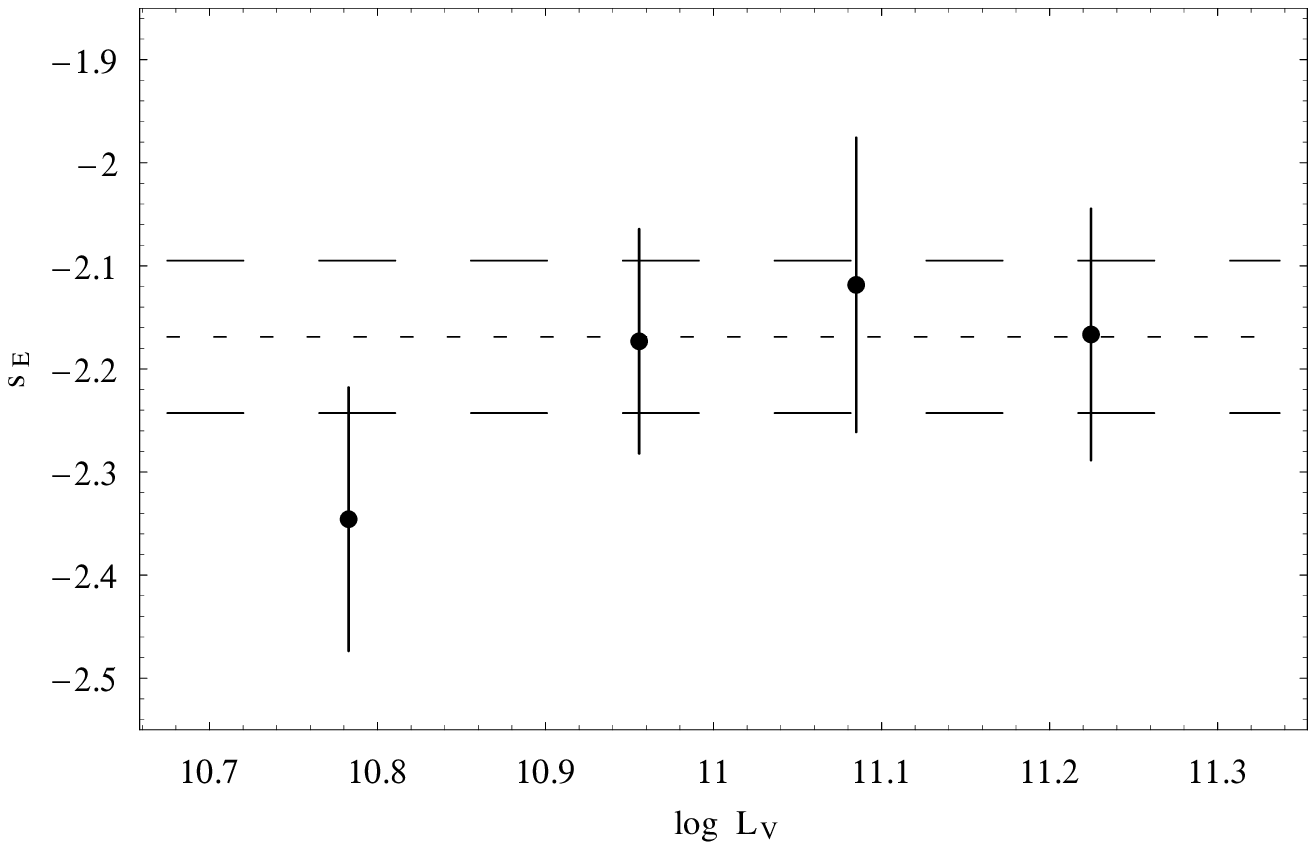}} \goodgap
\subfigure{\includegraphics[width=8cm]{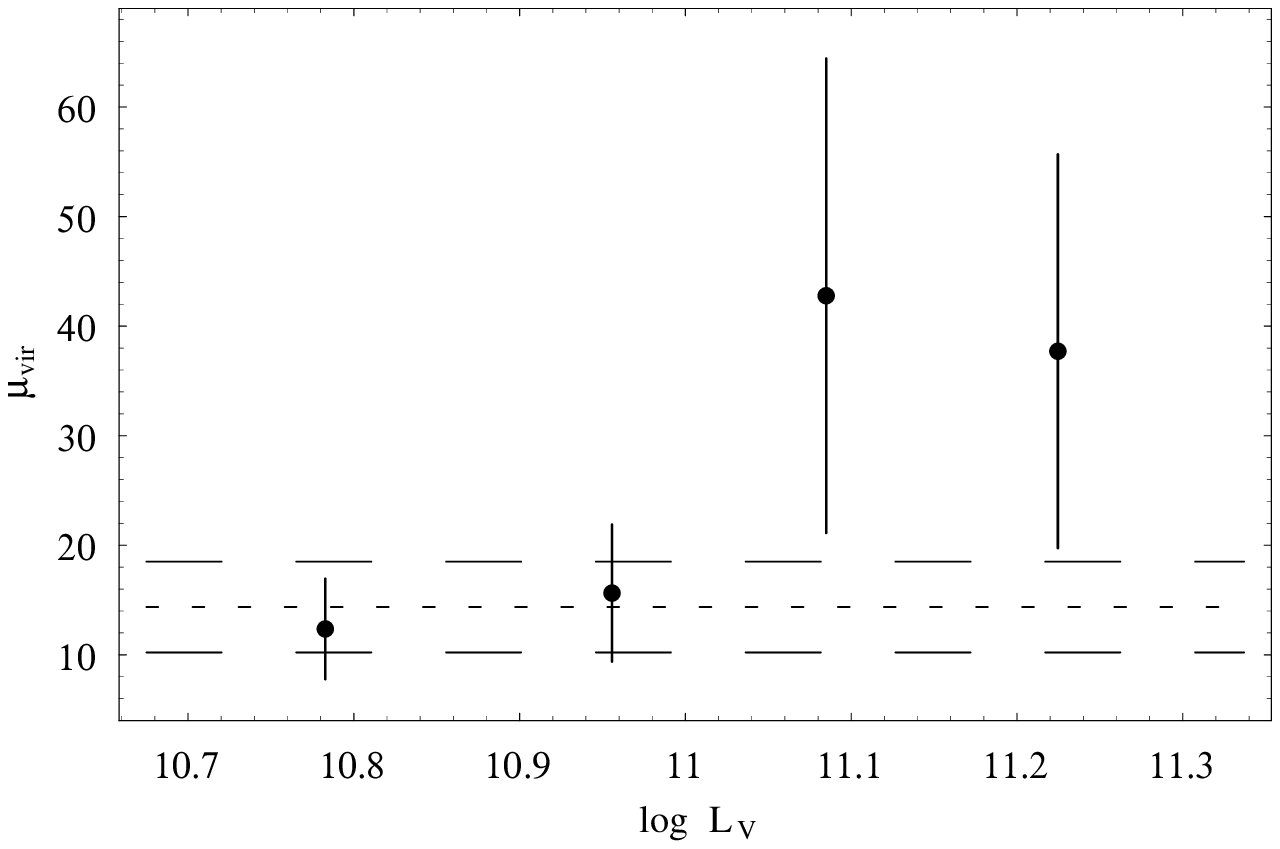}} \goodgap \\
\caption{Variation of the model parameters with the median $\log{L_V}$ of each bin.
The central dot denotes the median values, while the vertical bars span the $68\%$ CL range.
We overplot the constraints from the fit to the full sample as horizontal lines with
the short dashed one referring to the median value and the long dashed ones delimiting the $68\%$ CL range.}
\label{fig: varpar}
\end{figure*}

\begin{table*}
\begin{center}
\begin{tabular}{ccccc}
\hline
Sample & $\alpha$ & $\alpha + \beta$ & $s_E$ & $\mu_{vir}$ \\
\hline \hline

$~$ & $~$ & $~$ & $~$ & $~$ \\

B1 & $-0.11_{-0.14 \ -0.25}^{+0.16 \ +0.30}$ & $0.86_{-0.24 \ -0.41}^{+0.29 \ +0.58}$
& $-2.18_{-0.11 \ -0.23}^{+0.13 \ +0.25}$ & $27.01_{-12.62 \ -18.21}^{+23.32 \ +43.23}$ \\

$~$ & $~$ & $~$ & $~$ & $~$ \\

B2 & $-0.29_{-0.11 \ -0.15}^{+0.14 \ +0.23}$ & $0.91_{-0.21 \ -0.42}^{+0.20 \ +0.42}$
& $-2.12_{-0.14 \ -0.28}^{+0.14 \ +0.27}$ & $41.20_{-20.87 \ -32.06}^{+22.44 \ +32.07}$ \\

$~$ & $~$ & $~$ & $~$ & $~$ \\

B3 & $-0.28_{-0.11 \ -0.16}^{+0.10 \ +0.23}$ & $0.54_{-0.12 \ -0.21}^{+0.15 \ +0.31}$
& $-2.17_{-0.11 \ -0.19}^{+0.11 \ +0.21}$ & $12.02_{-4.44 \ -6.62}^{+8.06 \ +22.07}$ \\

$~$ & $~$ & $~$ & $~$ & $~$ \\

B4 & $-0.23_{-0.12 \ -0.20}^{+0.14 \ +0.19}$ & $0.51_{-0.12 \ -0.20}^{+0.22 \ +0.58}$
& $-2.34_{-0.13 \ -0.27}^{+0.12 \ +0.26}$ & $8.95_{-2.88 \ -3.81}^{+6.29 \ +21.25}$ \\

$~$ & $~$ & $~$ & $~$ & $~$ \\

\hline
\end{tabular}
\end{center}
\caption{Marginalized constraints on the model parameters for different bins.}
\end{table*}

Having established the validity of our model, let us now discuss
the constraints on its parameters. As a first lesson, we note that
models with $\alpha > 0$ are definitely excluded contrary to the
discussion in Sect.\,2 where we have shown that negative $\alpha$
values lead to a global \ML\ ratio diverging for $\eta \rightarrow
0$. However, such an unphysical behaviour is actually never
approached since, as we have seen in Sect.\,5, our
phenomenological ansatz do not fit realistic models for $\eta \le
0.01$ so that one has not to trust its extrapolation in the very
inner regions. On the other hand, the analysis in Sect.\,5 have
shown that models with $\alpha \le 0$ are needed to fit the global
\ML\ ratios of different dark halo models. It is therefore
reassuring that the likelihood analysis indeed prefer this kind of
models thus leading further support to our phenomenological
approach. As a further remark, we also note that the values of
$\alpha$ and $\alpha + \beta$ are typical of cored dark halo
density profiles (such as the NIS and Burkert models). However,
the very small $\alpha$ values obtained when considering the NFW
model are also partly due to the adopted mass\,-\,concentration
relation so that it is likely that changing the $c$\,-\,$M_{vir}$
relation may lead to larger and still negative $\alpha$ values.
Exploring this issue is outside our aims here, but we argue that
cored profiles are preferred on the basis of our phenomenological
approach.

A further success of our model is represented by the constraints
on $s_E$, i.e. the logarithmic slope of the total density profile
at the scaled Einstein radius, i.e. $s_E =
d\log{\rho}/d\log{\eta}(\eta = R_E/R_{eff})$. Previous analysis in
literature have typically fitted the observed Einstein radius and
velocity dispersion using a parameterized density profile. For
instance, Koopmans \& Treu (2003a,b) have fitted the data on
0047-281 and B1608+656 assuming $\rho \propto r^{-\gamma}$ and
finding $\gamma = -1.90^{+0.05}_{-0.23}$ and $\gamma = -2.03 \pm
0.14$, respectively. Using the same methodology, but averaging
over a sample of 15 SLACS lenses, Koopmans et al. (2006) have found again
$\gamma = -2.01^{+0.02}_{-0.03}$. As yet said before and also
stressed by the same authors, these may be
considered as constraints on the logarithmic slope at the Einstein
radius since this is the range mainly probed by the data. It is
therefore reassuring that, using a similar data analysis but a
radically different parametrization, we get compatible constraints
on $s_E$ thus strengthening our results.

Finally, we consider the $\mu_{vir}$ parameter. It is easy to
transform the constraints on this quantity into one on the DM mass
fraction at the virial radius thus getting\,:

\begin{displaymath}
\langle f_{vir} \rangle = (f_{vir})_{med} = 0.92 \ ,
\end{displaymath}

\begin{displaymath}
{\rm 68\% \ CL} \ : \ (0.89, 0.94) \ \ , \ \
{\rm 95\% \ CL} \ : \ (0.87, 0.96) \ \ .
\end{displaymath}
Using $f_{vir} = 1 - \varepsilon_{SF} f_b$ with $f_b \simeq 0.17$,
we get $\langle \varepsilon_{SF} \rangle \simeq 0.47$, i.e. only
$\sim 50\%$ of the baryonic mass is converted into stars. This is
consistent with  semianalytic galaxy formation models
(\cite{Benson2000}) which typically use values in the range $(0.2,
1.0)$. From the observational point of view, the situation is much
more complicated. On the one hand, studies matching the
luminosity and mass functions (\cite{MH02,GS02}) or $M/L$
gradients (\cite{Nap05}) of local galaxies suggest
$\varepsilon_{SF} \sim 0.25$, while stellar masses in the SDSS
survey (\cite{Pad04}) imply $\varepsilon_{SF} \simeq 0.2 - 0.8$.
Given such weak observational constraints, we can only argue that
our estimate of $\mu_{vir}$ and hence $\varepsilon_{SF}$ is fully
realistic. It is worth noting, however, that both observations and
theoretical modelling point towards a U\,-\,shaped trend of
$\varepsilon_{SF}$ with the mass (\cite{Benson2000,Nap05,conrwesh+08}).
One could therefore argue that our approach based on a universal
$\varepsilon_{SF}$ is unmotivated, but the small mass range probed by the lenses
and the good results obtained let us conclude that this choice
does not bias anyway the constraints on the quantities of interest.

\subsection{Binning the lenses}

A simple analysis of the best fit residual to the full sample
shows that there is actually no correlation with either the lens
redshift, the total luminosity and stellar mass. Such a result may
argue in favour of our assumption about the universality of the
parameters $(\alpha, \alpha + \beta, s_E, \mu_{vir})$. However,
this can also be a consequence of an erroneous estimation of the
errors or of insufficient statistics. To further
explore this issue, we therefore divide the lenses in four almost
equally populated bins (denoted as B1, B2, B3, B4) according to
their absolute $V$ magnitude and run our MCMC algorithm using
chains with $2 \times 10^{5}$ points reducing to 5601 after burn
in cut and thinning. Note that,  because we use now a smaller
number of constraints (namely, $6 - 8$ vs $38$), we have to run
longer chains in order to achieve the same convergence as with the
full sample. We summarise in Table 3 the best fit values 
and statistics of residuals for the fits to the binned samples, reporting 
in the second column the median value and the full range of $M_V$ in each bin,
while the two numbers in columns 4 and 5 refer to the mean and rms values of 
the lensing and dynamics residuals, respectively. Table 4 reports the constraints
on the model parameters after marginalization using the notation 
$x_{-y_1 \ -y_2}^{+z_1 \ +z_2}$ to mean that $x$ is the median value, 
while $68\%$ and $95\%$ CL read $(x - y_1, x + z_1)$ and $(x - y_2, x + z_2)$, 
respectively. These results allow us to make some interesting considerations.

As a first remark, we note that the fit is still successful as
witnessed by both the low normalized residuals and reduced
$\chi^2$ values\footnote{The high $\chi^2/d.o.f.$ for B2 is due to
to the peculiar lens SDSS J234111.6+000019 so remember the caveat
yet quoted before.}. Actually, one could wonder why the reduced
$\chi^2$ values for bin B1 and B4 are smaller than unity thus
arguing for a possible overestimate of the errors. While this is
possible since our estimate is based on a naive propagation not
taking into account correlations among the uncertainties on
photometric quantities, we nevertheless note that the
$\chi^2/d.o.f.$ values for bins B2 and B3 do not present such a
problem. Notwithstanding the solution of this ambiguity, we
however stress that the constraints on the model parameters are
not biased.

It is interesting to compare the results on the marginalized model
parameters with the corresponding ones for the full sample. To
this aim, we look at Fig.\,\ref{fig: varpar} where we have plotted
the median and $68\%$ CL estimate of each parameter as a function
of the median $\log{L_V}$ of the bin and overplotted the results
from the fit to the full sample. As it is clear, it is not
possible to infer any statistically reliable trend for $\alpha$
which is indeed constant within the errors and compatible with the
result from the fit to all lenses. A similar discussion may also
be done for $\alpha + \beta$, $\mu_{vir}$ and $s_E$, which due to
large uncertainties cannot allow to determine strong trends with
luminosity. However, these figures seem to suggest interesting
trends as a function of $\log{L_V}$, which we will better explore
in a future paper enlarging the sample. Our results give
issues about a dependence of the parameters above on luminosity
and stellar mass, with brighter galaxies having on average higher
values for $\alpha + \beta$ and $\mu_{vir}$ and a profile more
similar to the isothermal sphere. Finally, an increase of
$\mu_{vir}$ with luminosity and stellar mass implies that the star
formation efficiency $\varepsilon_{SF}$ decreases with these
quantities. Indeed, this is consistent with the observational and
theoretical studies (\cite{Benson2000,Nap05,conrwesh+08}, Tortora
etal. 2009a, Napolitano et al. 2009) showing that more massive
galaxies are less able to convert gas into stars.

\begin{table*}
\begin{center}
\begin{tabular}{ccccc}
\hline
Name & $f_{proj}(R_E)$ & $f_{DM}(R_E)$ & $f_{DM}(R_{eff})$ & $\nabla_{{\cal{l}}} \Upsilon$ \\
\hline \hline

$~$ & $~$ & $~$ & $~$ & $~$ \\

J002907.8-005550 & $0.615_{-0.016 \ -0.034}^{+0.016 \ +0.031}$ &
$0.564_{-0.016 \ -0.034}^{+0.016 \ +0.031}$
& $0.602_{-0.019 \ -0.041}^{+0.018 \ +0.033}$
& $0.47_{-0.12 \ -0.22}^{+0.13 \ +0.29}$ \\

$~$ & $~$ & $~$ & $~$ & $~$ \\

J015758.9-005626 & $0.660_{-0.019 \ -0.041}^{+0.018 \ +0.034}$ &
$0.601_{-0.016 \ -0.031}^{+0.014 \ +0.027}$
& $0.623_{-0.017 \ -0.036}^{+0.015 \ +0.029}$
& $0.56_{-0.14 \ -0.26}^{+0.15 \ +0.32}$ \\

$~$ & $~$ & $~$ & $~$ & $~$ \\

J021652.5-081345 & $0.715_{-0.013 \ -0.026}^{+0.012 \ +0.024}$ &
$0.687_{-0.011 \ -0.022}^{+0.010 \ +0.020}$
& $0.714_{-0.016 \ -0.035}^{+0.016 \ +0.031}$
& $0.62_{-0.19 \ -0.32}^{+0.20 \ +0.42}$ \\

$~$ & $~$ & $~$ & $~$ & $~$ \\

J025245.2+00358 & $0.666_{-0.016 \ -0.034}^{+0.016 \ +0.028}$ &
$0.621_{-0.013 \ -0.026}^{+0.011 \ +0.022}$
& $0.651_{-0.016 \ -0.035}^{+0.017 \ +0.030}$
& $0.55_{-0.15 \ -0.26}^{+0.16 \ +0.34}$ \\

$~$ & $~$ & $~$ & $~$ & $~$ \\

J033012.1-002052 & $0.631_{-0.021 \ -0.044}^{+0.020 \ +0.037}$ &
$0.558_{-0.018 \ -0.036}^{+0.017 \ +0.032}$
& $0.570_{-0.018 \ -0.037}^{+0.017 \ +0.032}$
& $0.53_{-0.13 \ -0.23}^{+0.15 \ +0.29}$ \\

$~$ & $~$ & $~$ & $~$ & $~$ \\

J072805.0+383526 & $0.563_{-0.021 \ -0.043}^{+0.019 \ +0.036}$ &
$0.474_{-0.018 \ -0.036}^{+0.024 \ +0.031}$
& $0.482_{-0.017 \ -0.035}^{+0.016 \ +0.031}$
& $0.46_{-0.11 \ -0.20}^{+0.11 \ +0.22}$ \\

$~$ & $~$ & $~$ & $~$ & $~$ \\

J080858.8+470639 & $0.609_{-0.019 \ -0.039}^{+0.017 \ +0.031}$ &
$0.543_{-0.016 \ -0.031}^{+0.013 \ +0.027}$
& $0.570_{-0.017 \ -0.037}^{+0.016 \ +0.030}$
& $0.48_{-0.12 \ -0.22}^{+0.13 \ +0.27}$ \\

$~$ & $~$ & $~$ & $~$ & $~$ \\

J090315.2+411609 & $0.678_{-0.023 \ -0.047}^{+0.021 \ +0.038}$ &
$0.615_{-0.019 \ -0.039}^{+0.019 \ +0.035}$
& $0.627_{-0.019 \ -0.039}^{+0.019 \ +0.035}$
& $0.61_{-0.15 \ -0.28}^{+0.16 \ +0.65}$ \\

$~$ & $~$ & $~$ & $~$ & $~$ \\

J091205.3+002901 & $0.730_{-0.012 \ -0.025}^{+0.015 \ +0.032}$ &
$0.714_{-0.015 \ -0.031}^{+0.018 \ +0.037}$
& $0.732_{-0.017 \ -0.034}^{+0.019 \ +0.039}$
& $0.61_{-0.20 \ -0.36}^{+0.22 \ +0.45}$ \\

$~$ & $~$ & $~$ & $~$ & $~$ \\

J095944.1+041017 & $0.575_{-0.021 \ -0.047}^{+0.019 \ +0.036}$ &
$0.523_{-0.026 \ -0.051}^{+0.023 \ +0.045}$
& $0.566_{-0.025 \ -0.055}^{+0.022 \ +0.041}$
& $0.42_{-0.11 \ -0.20}^{+0.12 \ +0.25}$ \\

$~$ & $~$ & $~$ & $~$ & $~$ \\

J102332.3+42300 & $0.546_{-0.020 \ -0.044}^{+0.020 \ +0.037}$ &
$0.454_{-0.018 \ -0.038}^{+0.018 \ +0.034}$
& $0.465_{-0.017 \ -0.038}^{+0.017 \ +0.033}$
& $0.44_{-0.11 \ -0.19}^{+0.11 \ +0.22}$ \\

$~$ & $~$ & $~$ & $~$ & $~$ \\

J110308.2+532228 & $0.705_{-0.017 \ -0.035}^{+0.023 \ +0.050}$ &
$0.692_{-0.023 \ -0.048}^{+0.028 \ +0.056}$
& $0.707_{-0.019 \ -0.041}^{+0.022 \ +0.048}$
& $0.54_{-0.18 \ -0.33}^{+0.19 \ +0.39}$ \\

$~$ & $~$ & $~$ & $~$ & $~$ \\

J120540.4+491029 & $0.622_{-0.016 \ -0.033}^{+0.016 \ +0.029}$ &
$0.573_{-0.016 \ -0.032}^{+0.015 \ +0.030}$
& $0.610_{-0.018 \ -0.040}^{+0.018 \ +0.033}$
& $0.48_{-0.13 \ -0.23}^{+0.13 \ +0.29}$ \\

$~$ & $~$ & $~$ & $~$ & $~$ \\

J125028.3+052349 & $0.592_{-0.018 \ -0.039}^{+0.018 \ +0.033}$ &
$0.527_{-0.017 \ -0.034}^{+0.015 \ +0.031}$
& $0.559_{-0.018 \ -0.040}^{+0.016 \ +0.034}$
& $0.46_{-0.11 \ -0.21}^{+0.12 \ +0.25}$ \\

$~$ & $~$ & $~$ & $~$ & $~$ \\

J140228.1+632133 & $0.654_{-0.014 \ -0.035}^{+0.016 \ +0.029}$ &
$0.605_{-0.013 \ -0.026}^{+0.012 \ +0.023}$
& $0.636_{-0.017 \ -0.036}^{+0.016 \ +0.030}$
& $0.53_{-0.14 \ -0.25}^{+0.15 \ +0.33}$ \\

$~$ & $~$ & $~$ & $~$ & $~$ \\

J142015.9+601915 & $0.489_{-0.043 \ -0.088}^{+0.033 \ +0.063}$ &
$0.437_{-0.056 \ -0.103}^{+0.045 \ +0.081}$
& $0.488_{-0.045 \ -0.099}^{+0.034 \ +0.064}$
& $0.34_{-0.09 \ -0.16}^{+0.09 \ +0.19}$ \\

$~$ & $~$ & $~$ & $~$ & $~$ \\

J162746.5-00535 & $0.666_{-0.015 \ -0.030}^{+0.013 \ +0.025}$ &
$0.626_{-0.013 \ -0.025}^{+0.012 \ +0.023}$
& $0.659_{-0.017 \ -0.035}^{+0.017 \ +0.031}$
& $0.54_{-0.16 \ -0.26}^{+0.16 \ +0.34}$ \\

$~$ & $~$ & $~$ & $~$ & $~$ \\

J163028.2+452036 & $0.639_{-0.022 \ -0.045}^{+0.020 \ +0.036}$ &
$0.568_{-0.018 \ -0.036}^{+0.017 \ +0.032}$
& $0.579_{-0.018 \ -0.037}^{+0.017 \ +0.031}$
& $0.55_{-0.14 \ -0.25}^{+0.14 \ +0.29}$ \\

$~$ & $~$ & $~$ & $~$ & $~$ \\

J230053.2+002238 & $0.660_{-0.016 \ -0.033}^{+0.015 \ +0.028}$ &
$0.615_{-0.014 \ -0.025}^{+0.011 \ +0.022}$
& $0.647_{-0.017 \ -0.036}^{+0.016 \ +0.030}$
& $0.54_{-0.15 \ -0.26}^{+0.15 \ +0.33}$ \\

$~$ & $~$ & $~$ & $~$ & $~$ \\

J230321.7+14221 & $0.646_{-0.015 \ -0.032}^{+0.013 \ +0.026}$ &
$0.608_{-0.016 \ -0.032}^{+0.015 \ +0.030}$
& $0.643_{-0.019 \ -0.038}^{+0.018 \ +0.033}$
& $0.50_{-0.14 \ -0.25}^{+0.15 \ +0.31}$ \\

$~$ & $~$ & $~$ & $~$ & $~$ \\

J234111.6+000019 & $0.659_{-0.014 \ -0.030}^{+0.017 \ +0.025}$ &
$0.627_{-0.016 \ -0.033}^{+0.014 \ +0.030}$
& $0.659_{-0.019 \ -0.038}^{+0.017 \ +0.034}$
& $0.51_{-0.15 \ -0.26}^{+0.16 \ +0.33}$ \\

$~$ & $~$ & $~$ & $~$ & $~$ \\

\hline
\end{tabular}
\end{center}
\caption{Median values and $68$ and $95\%$ CL ranges for the DM mass fractions
using the chain for the fit to the full lens sample.}
\end{table*}

\begin{figure*}
\centering
\subfigure{\includegraphics[width=8cm]{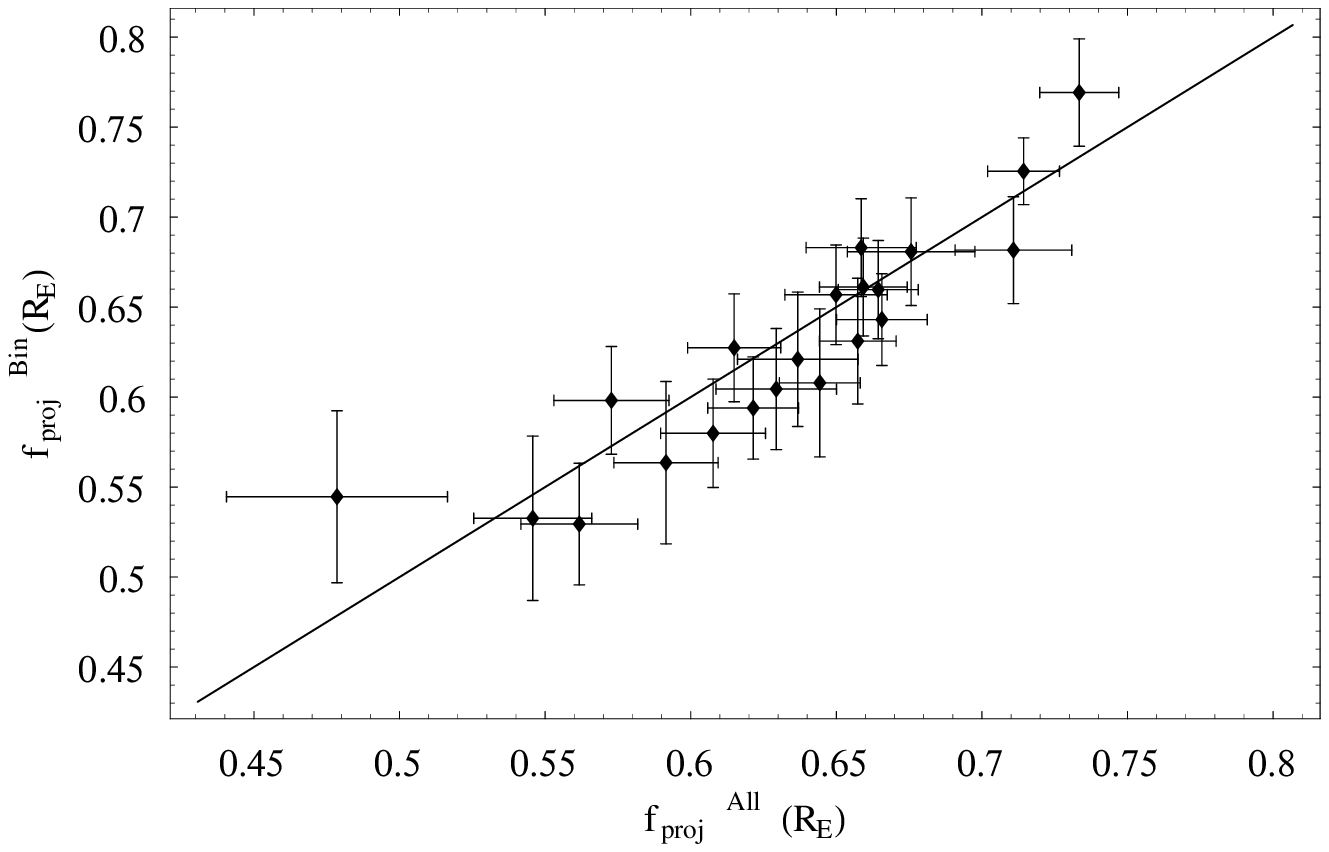}}
\goodgap \subfigure{\includegraphics[width=8cm]{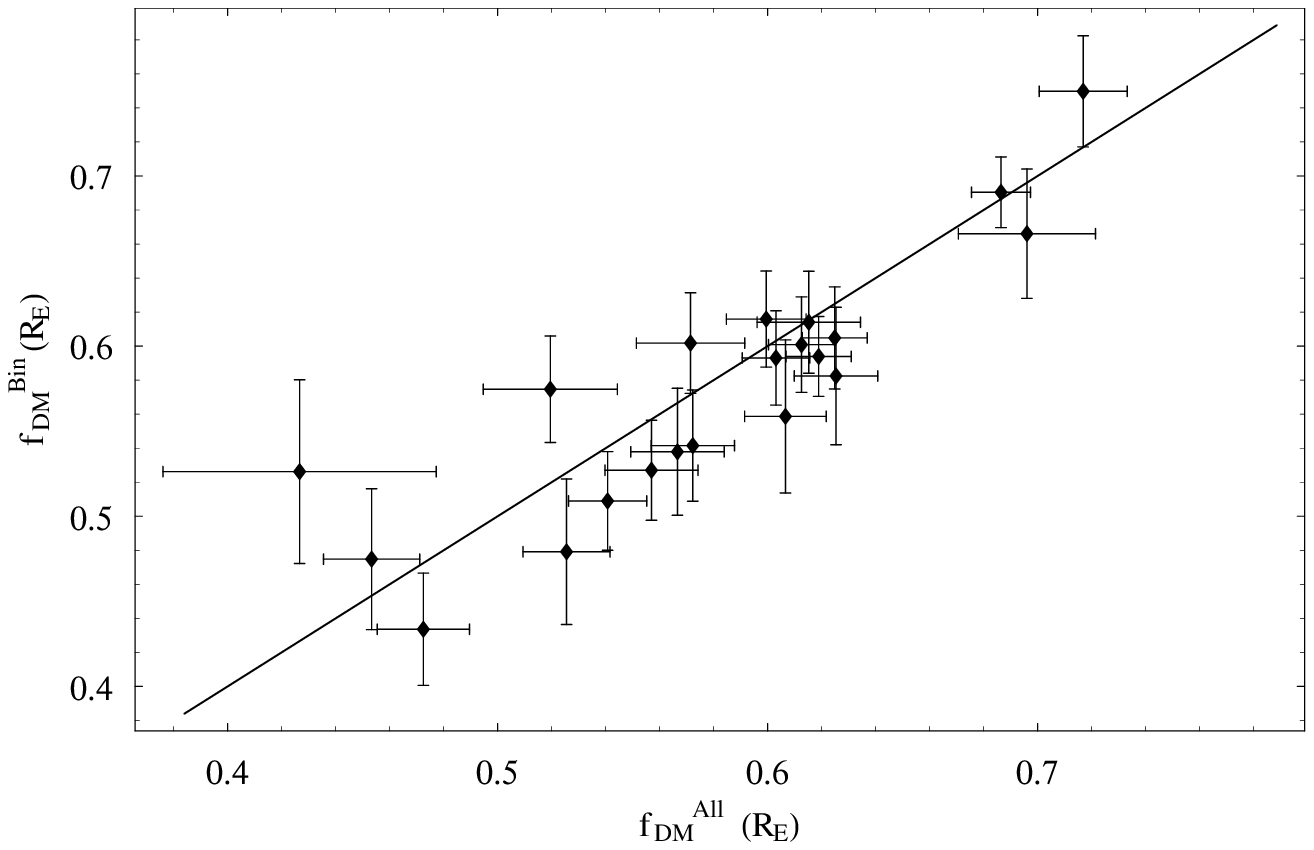}} \\
\subfigure{\includegraphics[width=8cm]{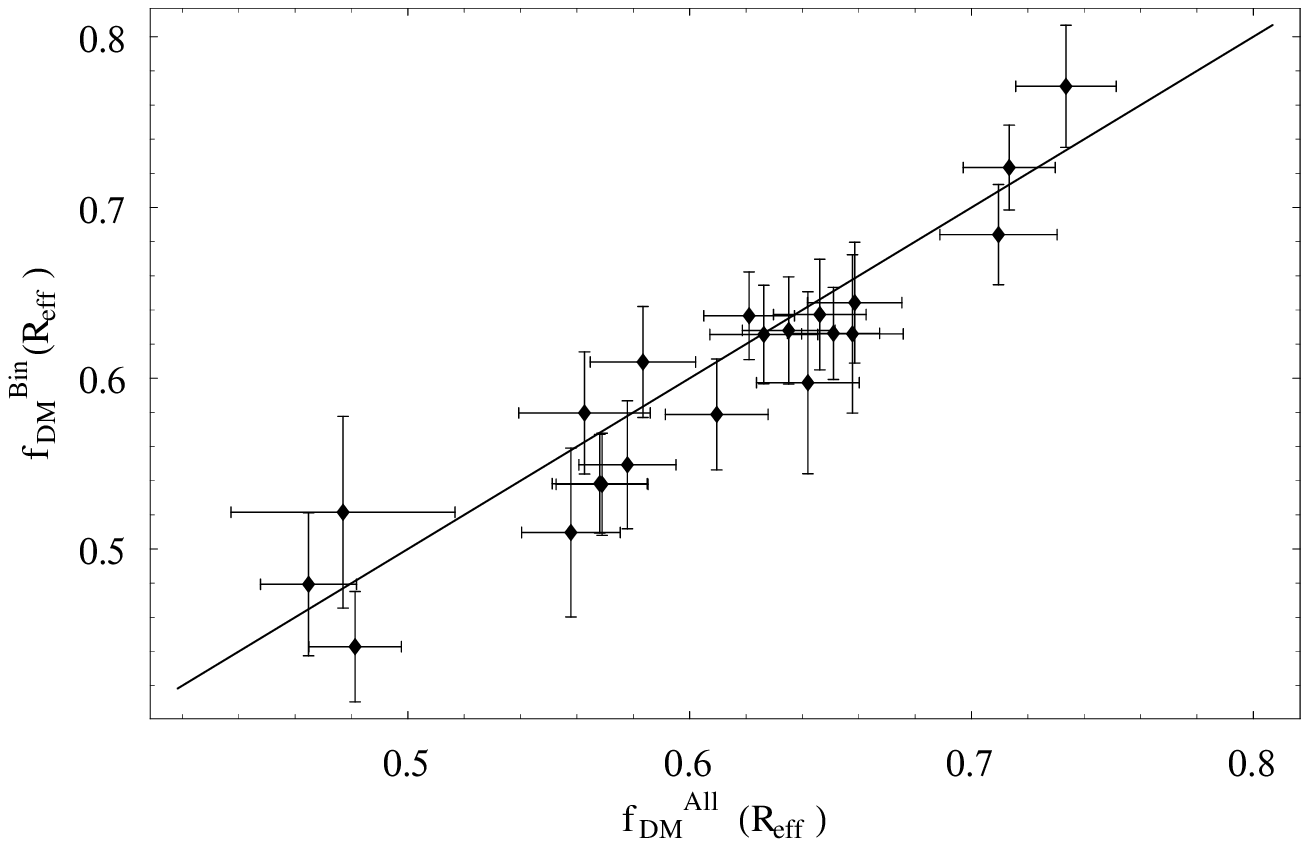}} \goodgap
\subfigure{\includegraphics[width=8cm]{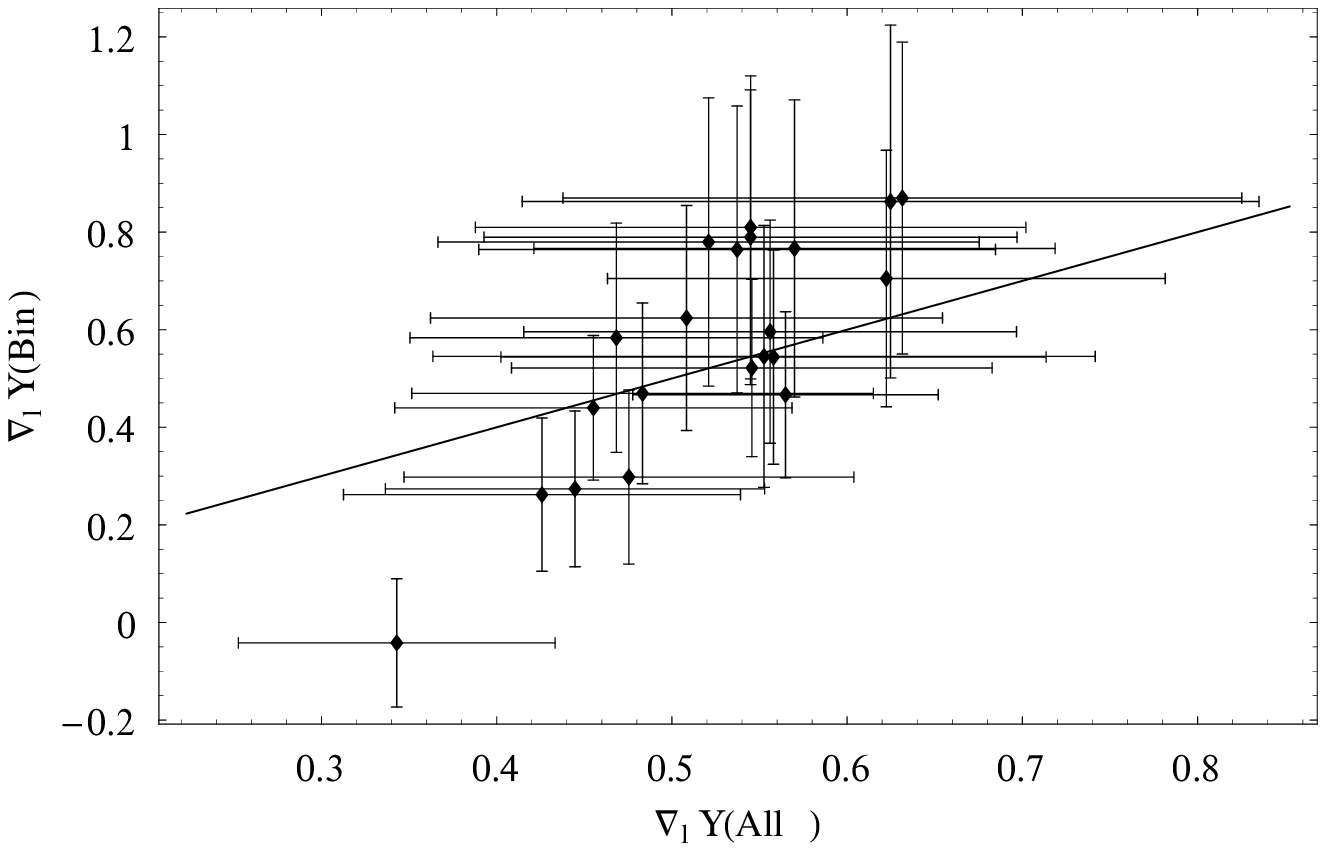}} \goodgap \\
\caption{Comparison between different DM content estimators as
constrained from the fit to the full or binned samples. Error bars
refer to the $68\%$ CL ranges. The solid line superimposed on the
data is the diagonal $f(x) = x$.} \label{fig: varfe}
\end{figure*}

As a general remark, it is worth noting that all the eventual
trends seen in Fig.\,\ref{fig: varpar} become statistically
meaningless should we have plotted the $95$ instead of the $68\%$
CL range. In contrast, one could equally well conclude that,
to a first reasonable approximation, the model parameters
$(\alpha, \alpha + \beta, s_E, \mu_{vir})$ do not depend on the
luminosity. As we will show later, the derived quantities (such as
the DM mass fractions and the scalelength parameter
$\log{\eta_0}$) are consistent within the bins and the full
sample. Motivated by these considerations, we therefore hereafter
refer to the constraints from the fit to the full lens sample. We,
however, plan to further explore this issue with a larger lens
sample covering a wider $M_V$ range in order to both improve
the statistics and increase the number of points to infer any
possible trend with the luminosity.

\subsection{The DM mass fraction}

Although the model we are testing has been introduced as an {\it
effective} one to describe the galaxy, it is nevertheless an ideal
tool to derive model independent constraints on the DM content of
ETGs. Indeed, Eq.(\ref{eq: upsfdm}) shows that constraining
$\Upsilon(r)$ is the same as constraining the spherical DM mass
fraction $f_{DM}(r)$. In order to quantify these results, we
summarize in Table 5, the median and $68$ and $95\%$ CL ranges for
the projected mass fraction $f_{proj}(R_E)$ at the Einstein radius
and for the spherial DM mass fraction at both the Einstein and
effective radii. To this end, we compute these quantities for each
point of the chain obtained from the fit to the full lens sample
after burn in cut and thinning.

Before discussing the constraints we obtain, it is worth wondering
whether they depend on the choice to fit the full lens sample or
the binned subsamples. To this end, we plot in Fig.\,\ref{fig:
varfe} the median values (with their $68\%$ CL ranges) for the
quantities of interest. It is worth noting that the errors bars
are actually underestimated since they do not take into account
the uncertainties on the stellar \ML\ ratio and the photometric
parameters $(I_e, R_{eff})$. Indeed, the typical uncertainty
increases in this case up to $\sim 0.1$ so that fully dominates
the error induced by the model parameters. We have, however,
adopted this choice in order to improve the reliability of the
figures. With this caveat in mind, Fig.\,\ref{fig: varfe} shows
that there is essentially a quite good agreement between the two
estimates. A direct fit\footnote{Both in Fig.\,\ref{fig:
varfe} and in the fits presented in this section, we first correct
for the small asymmetry in the errors following the prescriptions
given in D' Agostini (2004). Denoting with $x$ the central value
and with $\Delta_{\pm}$ the negative and positive error so that the
$68\%$ CL reads $(x - \Delta_{-}, x + \Delta_{+})$, the corrected
value is $x_{corr} = x + (\Delta_{+} - \Delta_{-})$ with a symmetric
uncertainty $\Delta_{corr} = (\Delta_{+} + \Delta_{-})/2$. Note that,
since the $68\%$ CL ranges in Table 5 are indeed quite symmetric, these
corrections are actually negligible for most of the lenses in our sample.}
(without taking errors into account) indeed gives\,:

\begin{displaymath}
f_{proj}^{Bin}(R_E) = 0.04 + 0.93 f_{proj}^{All}(R_E) \ \ (\sigma_{rms} = 0.02) \ \ ,
\end{displaymath}

\begin{displaymath}
f_{DM}^{Bin}(R_E) = 0.08 + 0.86 f_{DM}^{All}(R_E) \ \ (\sigma_{rms} = 0.04) \ \ ,
\end{displaymath}

\begin{displaymath}
f_{DM}^{Bin}(R_{eff}) = -0.003 + 0.99 f_{DM}^{All}(R_{eff}) \ \ (\sigma_{rms} = 0.03) \ \ .
\end{displaymath}
Considering that the typical uncertainties are much larger than
the offsets and that the slopes of these relations are close to 1,
we hereafter only considers the results from the fit to the full
sample as our estimates of the DM mass fractions\footnote{The
complete set of constraints, including the ones from the fit to
the binned samples, are available on request to the authors.}
summarized in Table 5.

\begin{figure*}
\centering
\subfigure{\includegraphics[width=8cm]{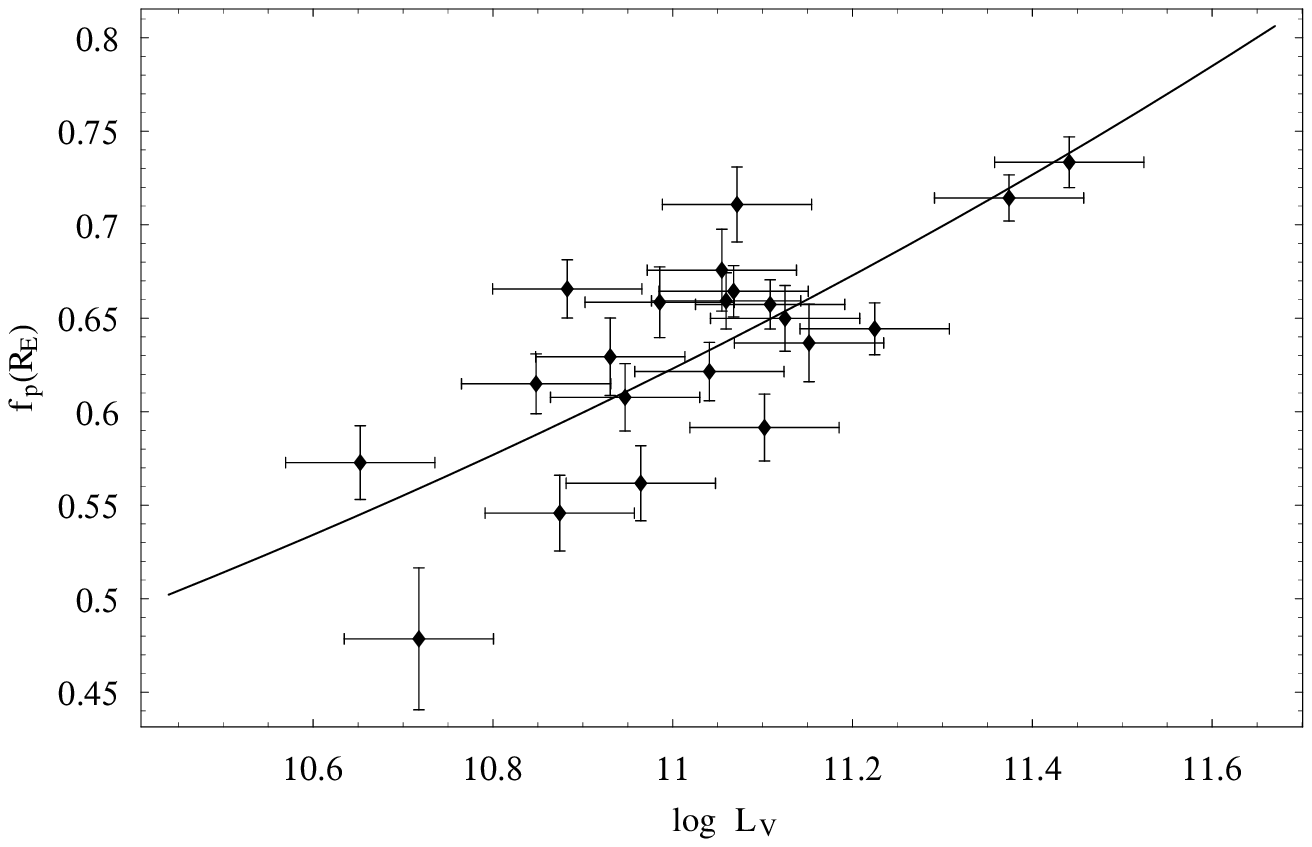}}
\goodgap \subfigure{\includegraphics[width=8cm]{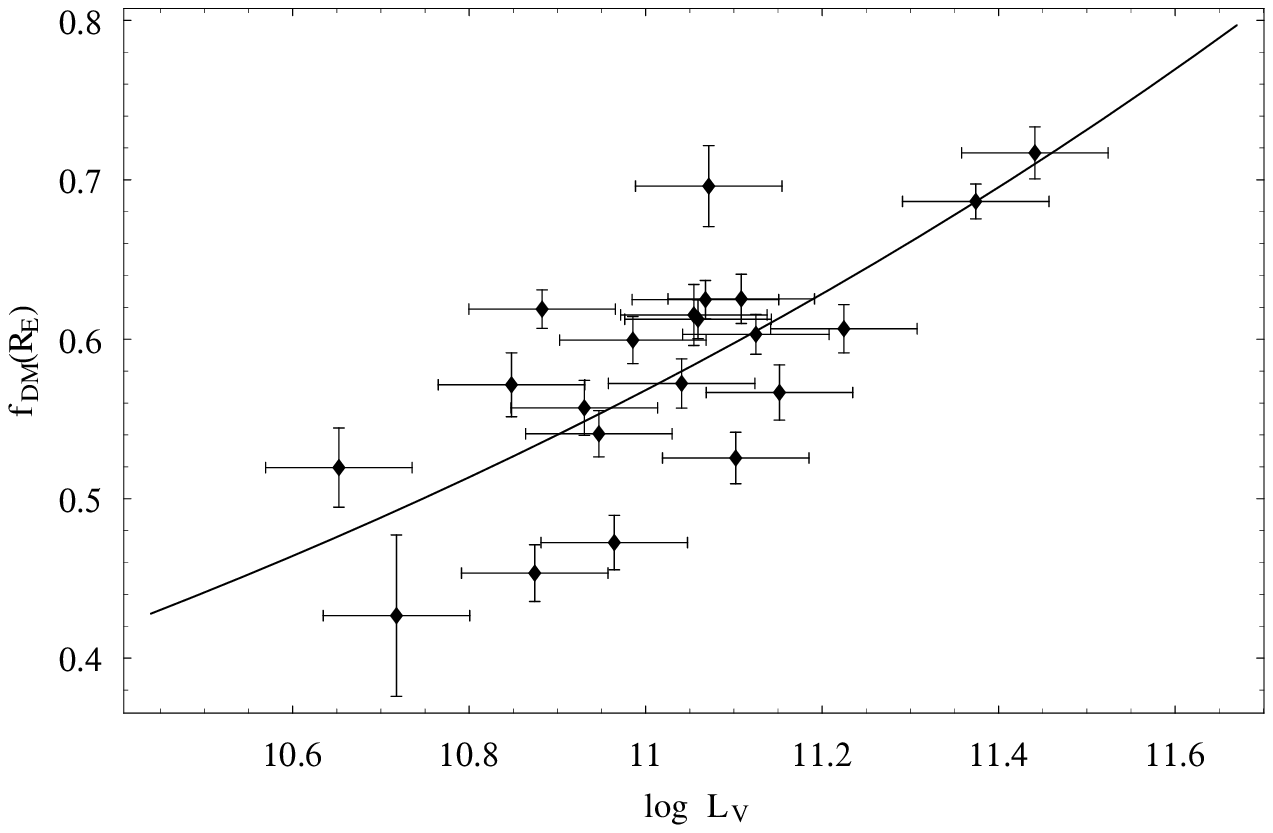}} \\
\subfigure{\includegraphics[width=8cm]{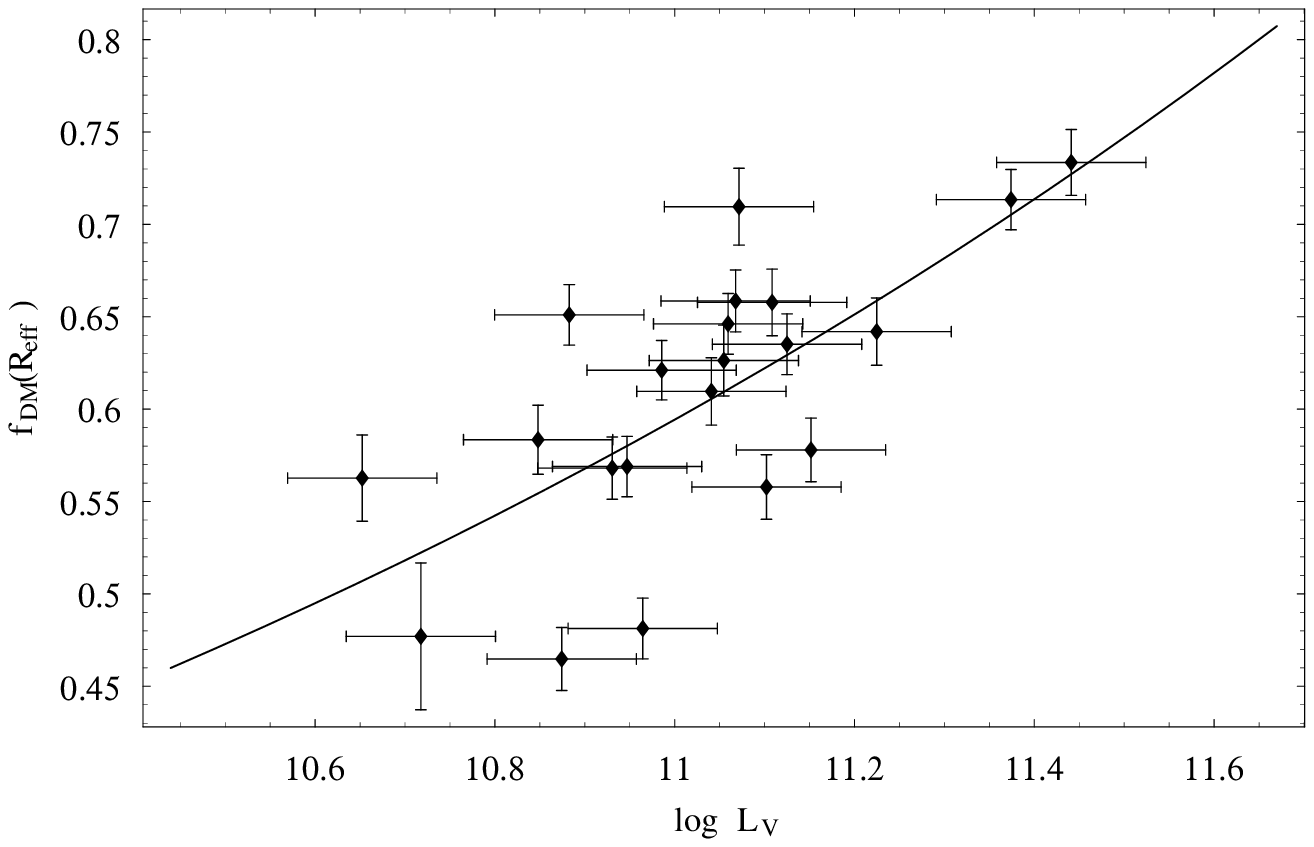}} \goodgap
\subfigure{\includegraphics[width=8cm]{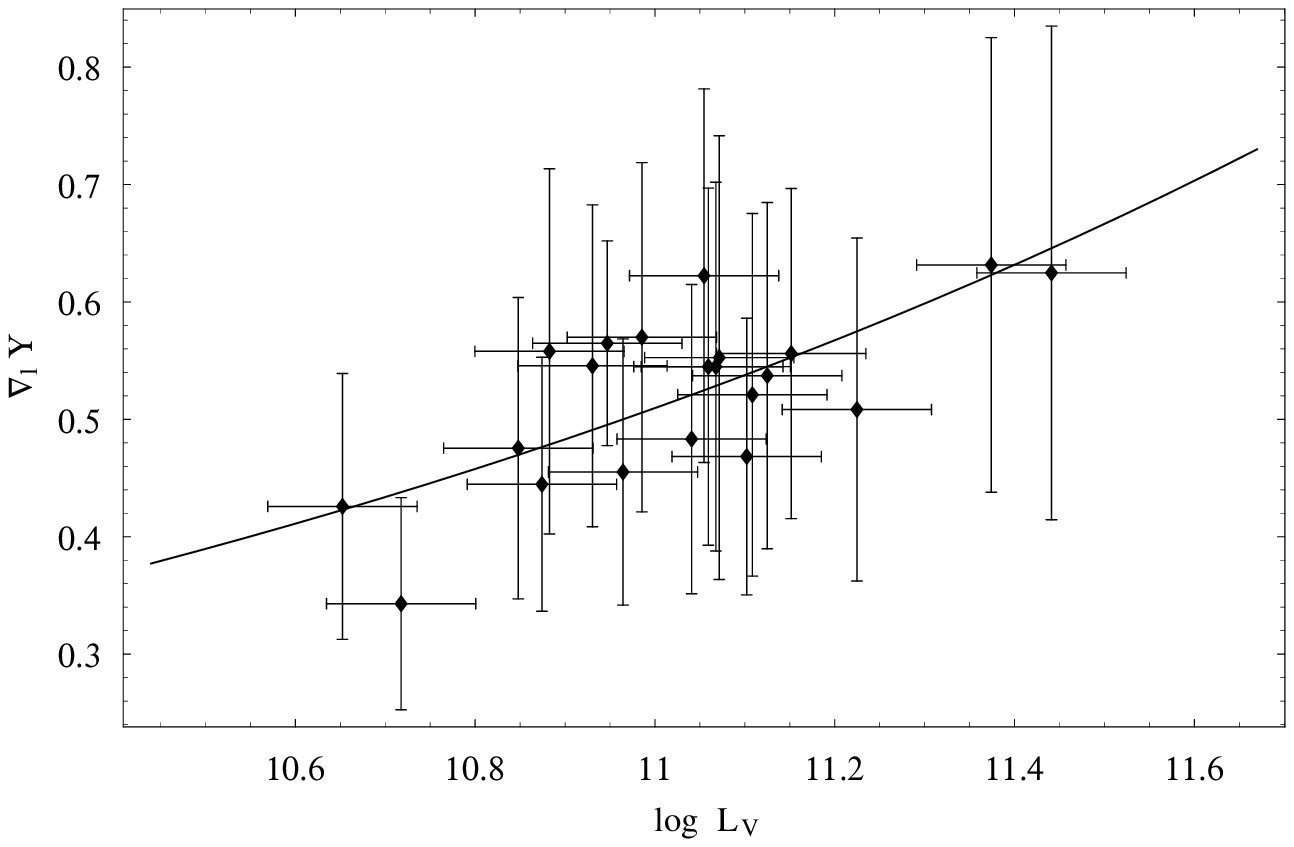}} \goodgap \\
\caption{DM mass fractions and empirical \ML\ gradient vs total
luminosity $\log{L_V}$. Errors bars refer to the $68\%$ CL
ranges, while the superimposed solid lines refer to the best fit
relations discussed in the text.} \label{fig: feplot}
\end{figure*}

We can further explore the viability of our results comparing our
estimated $f_{DM}(R_{eff})$ with previous estimates in literature.
Actually, a detailed comparison is not possible because our
systems are at an intermediate redshift ($z \simeq 0.1 - 0.5)$,
while dynamical and X\,-\,ray studies typically refer to nearby
galaxies. Moreover, such estimates are model dependent relying on
a given parametrization of the dark halo, while our constraints
come from a phenomenological and hence model independent analysis.
Since it is not possible to take quantitatively into account these
differences, we caution the reader to not overrate any agreement
or disagreement among our own and literature results.

As a first preliminary step, we briefly review some of the
previous determinations. An accurate quantitative estimate of
$f_{DM}(R_{eff})$ has been obtained by Gerhard et al. (2001) from
a dynamical analysis of the line profiles of 21 mostly luminous,
slowly rotating and nearly round ETGs with well measured velocity
dispersion and circular velocity profiles. According to their
preferred minimum halo models, the DM mass fraction is
$f_{DM}(R_{eff}) \simeq 10 - 40\%$ at the effective radius,
finding that their central dynamical \ML\ well agree with stellar
\ML\ estimated from synthetic models assuming a Salpeter IMF.
Using orbit superposition method, Thomas et al. (2005) managed to
reproduce in detail the observed kinematics of NGC 4807 finding
$f_{DM}(R_{eff}) = 0.21 {\pm} 0.14$ taking for \Yst\ a guess from
populations synthesis models with a Kroupa IMF. A larger sample of
24 ETGs with dynamics within $\sim 1 - 2 R_{eff}$ reconstructed
from integral field spectroscopy was considered by Cappellari et
al. (2006). Both two integrals analytic models and three integral
orbit superposition agree each other giving for $f_e$ a median
value $\simeq 0.3$ using a Kroupa IMF as input to the estimate of
\Yst. {Analyzing a sample of local galaxies, Tortora et al.
(2009a) found that luminous galaxies have typical values of
$f_{DM}(R_{eff}) \sim 0.3 - 0.4$, using a Salpeter IMF,
corresponding to $\sim 0.6 - 0.7$ for a Chabrier IMF.} Finally,
stacking the same 22 lenses we are using, Gavazzi et al. (2007)
used a combined weak and strong lensing analysis leaving the
stellar \ML\ as a free parameter of the fit, finding out
$f_{DM}(R_{eff}) = 0.27 {\pm} 0.04$. These authors estimate an
average value for stellar \ML\ of $\sim 3.1$ (using $h = 0.7$),
which is 1.5 times the median value we obtain using a Chabrier
IMF. Renormalizing all the estimates to the Chabrier IMF, the
estimates for $f_{DM}(R_{eff})$ then span the range $\sim (0.30,
0.70)$. Looking at Table 5, one can see that the values for
$f_{DM}(R_{eff})$ span a narrow range with $f_{DM}(R_{eff}) \simeq
0.61 \pm 0.07$ averaging over the 21 lenses in the sample. Such a
value is in agreement with local estimates even if it is worth
noting that most of the lenses have values of $f_{DM}(R_{eff})$ at
the upper end of the local range. Indeed, our estimates are on average
consistent with the local ones obtaibed in Tortora et al. (2009a)
when using the same Chabrier IMF adopted here. A more careful and
homogeneous (with respect to the local galaxies) sample is needed
to analyze the possible evolution of DM fraction
with redshift which is not the case here.

It has been recently reported in literature (Benson et al. 2000,
Marinoni \& Hudson 2002, Graham \& Guzman 2003, van den Bosch et
al. 2003a,b, Napolitano et al. 2005, Tortora et al. 2009a,b) that
the DM content correlates with the total luminosity, with more
luminous (and massive galaxies) having higher values of
$f_{DM}(R_{eff})$. This is what we indeed observe in
Fig.\,\ref{fig: feplot} where the projected and spherical DM mass
fractions (at both the Einstein and effective radii) are found to
correlate with $\log{L_V}$, with this latter quantity estimated at
the lens redshift. A direct fit, not taking into account
the errors, gives\,:

\begin{displaymath}
\log{f_{proj}(R_E)} = -2.04 + 0.17 \log{L_V} \ \ (\sigma_{rms} = 0.03) \ \ ,
\end{displaymath}

\begin{displaymath}
\log{f_{DM}(R_E)} = -2.66 + 0.22 \log{L_V} \ \ (\sigma_{rms} = 0.04) \ \ ,
\end{displaymath}

\begin{displaymath}
\log{f_{DM}(R_{eff})} = -2.41 + 0.20 \log{L_V} \ \ (\sigma_{rms} = 0.04) \ \ .
\end{displaymath}
Note that the low scatter is actually far smaller than the typical
uncertainty due to the errors on $\log{L_V}$ and \Yst, being of
order 0.1. The observed correlation with the luminosity suggests
that a similar correlations should be observed with the total
stellar mass. We indeed find\,:

\begin{displaymath}
\log{f_{proj}(R_E)} = -1.64 + 0.13 \log{M_{\star}} \ \ (\sigma_{rms} = 0.03) \ \ ,
\end{displaymath}

\begin{displaymath}
\log{f_{DM}(R_E)} = -1.98 + 0.15 \log{M_{\star}} \ \ (\sigma_{rms} = 0.05) \ \ ,
\end{displaymath}

\begin{displaymath}
\log{f_{DM}(R_{eff})} = -1.74 + 0.15 \log{M_{\star}} \ \ (\sigma_{rms} = 0.05) \ \ ,
\end{displaymath}
where $M_{\star}$ denotes here the total stellar mass. Although
encouraging, such correlations should be taken with some caution
because of the large measurement errors on the mass fractions and
the relatively narrow luminosity and mass range probed (spanning
roughly only one order of magnitude). It is therefore mandatory to
both increase the sample and reduce the errors on the estimated
stellar \ML\ ratio before trying any detailed comparisons with
available theoretical scenarios.

A different way of characterizing the galaxy DM content has been
proposed by Napolitano et al. (2005) introducing the empirical DM
gradient\,:

\begin{displaymath}
\nabla_l \Upsilon =
\frac{R_{eff}}{r_{out} - r_{in}} \left [ \left .
\frac{M_{DM}(r)}{M_{\star}(r)} \right |_{r = r_{out}} -
 \left . \frac{M_{DM}(r)}{M_{\star}(r)} \right |_{r = r_{in}} \right ]
\end{displaymath}
which, for our model, reduces to\,:

\begin{displaymath}
\nabla_l \Upsilon = \frac{(\eta_{out} + \eta_0)^{\beta} \eta_{out}^{\alpha} -
(\eta_{in} + \eta_0)^{\beta} \eta_{in}^{\alpha}}
{[1 - f_{DM}(R_{eff})](\eta_{out} - \eta_{in}) (1 + \eta_0)^{\beta}} \ .
\end{displaymath}
We estimate this quantity for the lenses in our sample using the
constraints from both the fit to the full lens sample and to the
binned samples setting ($\eta_{in}, \eta_{out}) = (0.5, 4.0)$ as
in Napolitano et al. (2005). A comparison between the two is shown
in the lower right panel in Fig.\,\ref{fig: varfe}. Although the
two set of constraints are in clear disagreement if we consider
only the median values (with a direct fit giving $\nabla_l
\Upsilon(Bin) \propto 0.83 \nabla_l \Upsilon(All)$ with a
significant scatter), the error bars are so large that we can
neglect this discrepancy and rely on the results of the fit to the
full sample. The lower right panel in Fig.\,\ref{fig: feplot} then
shows that this quantity is found to correlate with the
luminosity. It is worth mentioning this trend with luminosity will
be steeper if considering the fit binning the lenses, in agreement
with a star formation efficiency decreasing with luminosity and
mass (\cite{Nap05}). Using, however, the estimates from the fit to
the full lens sample, we find\,:

\begin{displaymath}
\log{\nabla_l \Upsilon} = -2.86 + 0.23 \log{L_V} \ \ (\sigma_{rms} = 0.04) \ \ ,
\end{displaymath}

\begin{displaymath}
\log{\nabla_l \Upsilon} = -2.53 + 0.20 \log{M_{\star}} \ \ (\sigma_{rms} = 0.04) \ \ .
\end{displaymath}
Fig.\,6 of Napolitano et al. (2005) plots the estimated $\nabla_l
\Upsilon$ values of a sample of local ETGs vs the total stellar
mass, with $9.5 \le \log M_{\star} \le 12$. The empirical \ML\
gradient seems to be a gently falling function of $\log M_{\star}$
over the range $9.5 \le \log M_{\star} \le 10.5$ to become then an
increasing function with a slope depending on the mass range
considered. Indeed, over the mass range probed by our sample, the
increase is quite gently (if not constant) so that our result
$\nabla_l \Upsilon \propto M_{\star}^{0.20}$ is in qualitative
good agreement with what may be inferred by the local sample.
It is nevertheless worth stressing that our $\nabla_l \Upsilon$
values are consistent with those in Napolitano et al. (2005).
Needless to say, a larger statistics is mandatory in oder to
confirm these results and compare them with the trend predicted on
the basis of dark halo models.

\subsection{The impact of the IMF choice}

All the results discussed insofar rely on the initial choice of
the stellar IMF. Adopting a different IMF, indeed, changes the
estimated stellar \ML\ ratios \Yst\ and hence the values of
$M_{vir}$ and then of $\Upsilon_{eff}$ through Eq.(\ref{eq:
upseffmvir}). It is therefore worth wondering how the results
change with the IMF. To this aim, we rescale all the stellar \ML\
ratios by a factor 1.8, that is the same as assuming a Salpeter
rather than a Chabrier IMF. Then we repeat the fit to the full
sample with these rescaled \Yst\ and compare the best fit
parameters and the marginalized constraints with those in Table 2.

Not surprisingly, we find a very good agreement for $(\alpha,
\alpha + \beta, s_E)$, while the only parameter changed is
$\mu_{vir}$. This can be easily understood considering that, for a
given lens, rescaling \Yst\ is the same as rescaling
$\Upsilon_{eff}$ leaving unaltered the shape of the global \ML\
ratio $\Upsilon(r)$. In order to still fit the data, we have
therefore to retain the same functional expression and hence the
same values of the slope parameters $(\alpha, \alpha + \beta,
s_E)$, but scale $\Upsilon_{eff}$ of the same amount as the
stellar \ML\ ratios. Since the luminosity is the same and the
virial mass has to remain unchanged in order to give the same
values of the projected mass and velocity dispersion, the virial
DM mass fraction has to be rescaled according to\,,:

\begin{displaymath}
f_{vir}^S =  1 - \Yst^S L_T/M_{vir} = 1 - (1 - f_{vir}^C)
(\Yst^S/\Yst^C)
\end{displaymath}
and for the mass scaling parameter we get\,:

\begin{displaymath}
\frac{\mu_{vir}^S}{\mu_{vir}^C} = \left ( \frac{M_{vir}}{\Yst^S
L_T} \right ) \left ( \frac{M_{vir}}{\Yst^S L_T} \right )^{-1} =
\frac{\Yst^C}{\Yst^S} \simeq 0.55
\end{displaymath}
in agreement with what we find within the errors.

As a consequence, all the results obtained on the correlations
between the DM mass fractions and the stellar luminosity and mass
are still valid. Indeed, to take account of the change in the IMF,
we have simply to rescale the $f_{DM}$ values using $f_{DM}^S =
1 - 1.8 (1 - f_{DM}^C)$, while $\log{M_{\star}}$ has
to be increased by a constant factor $\simeq 0.25$. Both
these changes obviously rescales the corresponding zeropoints and
make the slope analyzed steeper than those obtained using a
Chabrier IMF strongly pointing to an increasing DM content as a
function of luminosity and stellar mass.

\section{Conclusions}\label{sec:conclusions}

According to the estimated baryon and total matter density
parameter predicted by the cosmological concordance model, DM
should represent more than $80\%$ of the total matter budget. As
such, it is supposed to be ubiquitous representing therefore a
significant component of any galactic structure.  Notwithstanding
this reasonable hypothesis, there is still a strong debate
regarding the DM content of ETGs with a variety of dynamical,
lensing and X\,-\,ray studies leading to contrasting results. The
lack of a reliable mass tracer in the ETGs outer regions and
modeling uncertainties make the problem of DM in ETGs a difficult
one to conclusively address. In order to reduce any possible
systematic bias due to the choice of the dark halo profile, a
phenomenological approach has been proposed in Tortora et al.
(2007) relying on the use of a versatile analytical expression for
the global \ML\ ratio $\Upsilon(r)$. Such a method allows to
smoothly interpolate between the two extreme cases of {\it light
traces mass} and dark halo dominance mimicking well the main
dynamical properties of a large class of intermediate models. As a
next mandatory step, we have here investigated the viability of
the model by contrasting it against the measured Einstein angles
and aperture velocity dispersions of a sample of 21 lenses
observed by the SLACS survey. A  Bayesian statistical analysis
makes it possible to obtain the main results sketched below.

\begin{enumerate}

\item{The {\it effective} galaxy model obtained by combining our ansatz for
$\Upsilon(r)$ with the deprojection of the Sersic profile provided
by the Prugniel \& Simien (1997) approximate expression is able to
reproduce the observations under the assumptions that the slope
parameters $(\alpha, \alpha + \beta)$, the logarithmic slope of
the total density profile at the scaled Einstein radius, $s_E =
d\ln{\rho}/d\ln{\eta}(\eta = R_E/R_{eff})$, and the virial mass ratio,
$\mu_{vir} = M_{vir}/\Yst\ L_T$, are universal quantities, while the
model scalelelength $\eta_0 = r_0/R_{eff}$ and global \ML\ ratio at
the effective radius $\Upsilon_{eff}$ may change on a case\,-\,by\,-\,case basis. \\}

\item{According to the marginalized constraints from the fit to the full
lens sample, the slope parameters $\alpha$ and $\alpha + \beta$ are in agreement
with what is expected in order the model mimics well cored halo profiles, while the
total density profile is locally approximated by an isothermal model as yet found
by previous studies in literature. We therefore argue that previous finding that an
isothermal model fits the lensing and dynamics data is likely a consequence of
having forced the total density profile to have a constant slope or of the data
covering a too limited radial range to detect the actual change in the logarithimic slope. \\}

\item{Binning the lenses according to their absolute $V$ magnitude does not
improve the fit and, on the contrary, gives constraints on both
the model parameters and DM mass fractions that are in agreement
(within $68$ or $95\%$ CL ranges) with those coming out from the
fit to the full lens sample. However, the small statistics and the
limited luminosity range probed does not allow us to draw any
definitive conclusion on the universality of the proposed
phenomenological $\Upsilon(r)$ model. \\}

\item{Using stellar population synthesis models and a Chabrier IMF
to estimate the stellar \ML\ ratios, we can constrain both the
projected and spherical DM mass fractions at the Einstein and
effective radii. Moreover, we have also estimated the empirical
$M/L$ gradients proposed in Napolitano et al. (2005) as a
different way to quantify the DM content in ETGs. \\}

\item{There is a clear trend of $f_{DM}(R_{eff})$ with both $\log{L_V}$
and $\log{M_{\star}}$ showing that the more luminous and/or massive a galaxy is, the
larger is its DM content within the effective radius. Unfortunately, the ranges in
$\log{L_V}$ and $\log{M_{\star}}$ probed are too small to make any quantitative
comparison with theoretical models. \\}

\end{enumerate}
Although quite interesting and in good agreement with previous
findings, the above results are still affected by too large
uncertainties so that it is worth wondering how they can be
strengthened. On one hand, one can simply increase the lens
sample. Indeed, while this work was nearly at its end, the SLACS
collaboration released a final catalog (\cite{B08}) containing 170
lens candidates from which a subsample of 53 lenses with reliable
determinations of the photometry, Einstein angle and aperture
velocity dispersion. One can naively expect that more than
doubling the number of objects will make our analysis more
stringent making it possible to increase both the number of bins
and the objects within a given bin. Unfortunately, more statistics
does not automatically imply higher precision. This is likely the
case for our model since the data are still probing only a limited
range centred on $0.5 - 0.7 \ R_{eff}$ so that stacking more
lenses does not increase the radial range probed. As a
consequence, therefore, we do expect to narrow the constraints on
the parameters mainly determing the shape of $\Upsilon(r)$ over
this range (i.e., $\alpha$ and $s_E$), while we cannot anticipate
whether a similar improvement may be achieved for $\alpha + \beta$
and $\mu_{vir}$ and hence $f_{DM}(R_{eff})$. Therefore, rather
than reducing the error on this quantity (that also depends on
that on \Yst\ and the photometric parameters $I_e$ and $R_{eff}$),
we argue that such a worth to do analysis will improve the
study of the $f_{DM}(R_{eff})$\,-\,${L_V}$ and
$f_{DM}(R_{eff})$\,-\,${M_{\star}}$ relations thus improving the
comparison with previous results and theoretical expectations. It
is worth noting, however, that the SLACS survey selection function
is biased against low mass systems (\cite{B06}) so that such an
analysis will likely be unable to explore the full ETGs mass range
which could lessen the comparison with galaxy formation scenarios.
Enlarging the sample will allow the possible physical interpretations  
of the trend we obtain to be better investigated. The change of DM
within effective radius as a function of mass (luminosity) scale
is the main driver of Fundamental Plane (\cite{cresc08a}), due to
the change of effective radius (\cite{shabern+09}) and/or star
formation efficiency (\cite{Nap05},2009). Galaxy merging, close
encounter, supernovae and AGN feedback could be the physical
phenomena which give reason of these trends (Benson et al. 2000,
Kaviraj et al. 2007, Khalatyan et al. 2008, Schawinski et al.
2008, Ruszkowski \& Springel 2009, \cite{shabern+09},
\cite{Tortora09AGN}).

As a complementary approach, one can resort to a different class
of objects. Rather than concentrating on intermediate redshift
galaxies, nearby ETGs may be investigated relying on the velocity
dispersion profile to probe the model in the inner regions, while
planetary nebulae may extend deep in the outer regions thus
constraining  the outer asymptotic slope. The analysis of the best
fit residuals here obtained has shown that no correlation with the
redshift is present thus suggesting that the model parameters
$(\alpha, \alpha + \beta, s_E, \mu_{vir})$ do not evolve. Such a
conclusion must be, however, checked by both dividing lenses in
redshift bins and looking at a nearby sample. Should this result
be confirmed, one could then investigate what physical interplay
between stellar and dark components is able to give rise to a
global \ML\ ratio whose shape does not evolve with $z$. In
future analysis we will analyze this problem.

As a final comment, we would like to stress the virtue of a
phenomenological approach (not necessarily our own) to the problem
of DM in ETGs. A comparison with  the determination of the dark
energy equation of state $w(z)$ is worth doing here. Since  a
plethora of theoretically motivated models provide different
expressions for $w(z)$, it has soon become clear that fitting all
of them to the different data available  would be a prohibitive
task. This lead to the emergence of phenomenological proposals for
$w(z)$ able to explore a wide range of models in a single step
also providing model independent constraints to scrutinize
different theories. Although the data on ETGs lensing and dynamics
are still limited, it is nevertheless yet possible to rely  on
that positive experience to efficiently tackling the problem of DM
in ETGs in a model independent way.

\begin{acknowledgements}
We warmly thank the referee for his/her constructive report which
significantly helped us to improve the paper. VFC is supported by 
University of Torino and Regione Piemonte and partially from the 
INFN project PD51. CT is funded by a grant from the 
project Mecenas funded by the Compagnia di San Paolo.
\end{acknowledgements}

\end{document}